\documentclass[useAMS,usenatbib]{mn2e}

\usepackage{graphicx}
\usepackage{amsmath}
\usepackage{amssymb}

\title[Mass accretion vs stellar age]{Strong Biases in Estimating the Time Dependence of Mass Accretion Rates in Young Stars}

\author[N. Da Rio et al.]
  {N.~Da Rio,$^{1,2}$\thanks{ndario@ufl.edu}
   R.~D.~Jeffries,$^3$
   C.~F.~Manara,$^4$
   M.~Robberto,$^5$
  \\
  $^1$Department of Astronomy, University of Florida, 211 Bryant Space Science Center, Gainesville, FL 32611, USA\\
  $^2$European Space Agency, Keplerlaan 1, 2200 AG Noordwijk, The Netherlands\\
  $^3$Astrophysics Group, Research Institute for the Environment, Physical Sciences and Applied Mathematics, Keele University, \\ Staffordshire ST5 5BG\\
  $^4$European Southern Observatory, Karl Schwarzschild-Str. 2, D-85748 Garching, Germany\\
  $^5$Space Telescope Science Institute, 3700 San Martin Dr., Baltimore MD, 21218, USA\\
}
\begin{document}

\date{MNRAS in prep}

\pagerange{\pageref{firstpage}--\pageref{lastpage}} \pubyear{2014}

\maketitle

\label{firstpage}

%%%%%%%%%%%%%%%%%%%%%%%%%%%%%%%%%%%%%%%%%%%%%%%%%%%%%%%%%%%%%%%%%%%%%%%%%%%%%%%%%%

\begin{abstract}
The temporal decay of mass accretion in young stars is a fundamental tracer of the early evolution of circumstellar disks. Through population syntheses, we study how correlated uncertainties between the estimated parameters of young stars (luminosity, temperature, mass, age) and mass accretion rates $\dot M_{\rm acc}$, as well as observational selection effects, can bias the temporal decay of mass accretion rates ($\dot M_{\rm acc}\propto t^{-\eta}$) inferred from a comparison of measured $\dot M_{\rm acc}$ with isochronal ages in young stellar clusters.

We find that the presence of realistic uncertainties reduces the
measured value of $\eta$ by up to a factor of 3, leading to the
inference of shallower decays than the true value. This suggests a much
faster temporal decay of $\dot M_{\rm acc}$ than generally
assumed. When considering the minimum uncertainties in ages affecting
the Orion Nebula Cluster, the observed value $\eta\sim1.4$, typical of
Galactic star forming regions, can only be reproduced if the real decay
exponent is $\eta\gtrsim 4$. This effect becomes more severe if one
assumes that observational uncertainties are larger, as required by
some fast star formation scenarios. Our analysis shows that while
selection effects due to sample incompleteness do bias $\eta$, they can
not alter this main result and strengthen it in many cases. A remaining
uncertainty in our work is that it applies to the most commonly
used and simple relationship between $\dot M_{\rm acc}$, the
accretion luminosity and the stellar parameters. We briefly explore how
a more complex interplay between these quantities might change the results.

\end{abstract}

%%%%%%%%%%%%%%%%%%%%%%%%%%%%%%%%%%%%%%%%%%%%%%%%%%%%%%%%%%%%%%%%%%%%%%%%%%%%%%%%%%


%%%%%%%%%%%%%%%%%%%%%%%%%%%%%%%%%%%%%%%%%%%%%%%%%%%%%%%%%%%%%%%%%%%%%%%%%%%%%%%%%%

\section{Introduction}
\label{section:intro}
Accretion of material from the circumstellar disk onto the central star is a common process occurring during the pre-main sequence (PMS) phase of stellar evolution \citep{hartmann1998}. Disk accretion is responsible for the build up of a significant fraction of the final stellar mass, thus affecting in part the shape of the initial mass function (IMF). On-going accretion is also direct evidence of the presence of gaseous circumstellar disc material close to the star and therefore the time dependence of the mass accretion rate ($\dot{M}_{\rm acc}$) traces disc evolution. Moreover, accretion influences the disk structure and thus contributes to the conditions for planet formation. Understanding the temporal evolution of mass accretion rates is crucial for modelling both planet formation and the early evolution of stars.

The dependence of $\dot M_{\rm acc}$ on age $t$ and stellar mass $M_*$ has been investigated both theoretically and observationally \citep[e.g.,][]{hartmann1998,muzerolle2003,muzerolle2005,robberto2004,sicilia-aguilar2006,sicilia-aguilar2010,antoniucci2011,biazzo2012}. A general finding is that $\dot M_{\rm acc}$ decreases with time -- indicating that the inner circumstellar disk is cleared within a few Myr of PMS evolution -- and scales with stellar mass. The large scatter in the data does not allow an accurate constraint on the form of this relation, which is commonly assumed to be a power law:
\begin{equation}
\label{equation:mdot_propto}
\dot M_{\rm acc}\propto t^{-\eta} M_*^b.
\end{equation}
Measurements in a number of nearby star forming regions generally find $\eta\sim 1$-$1.5$ and $b\sim 1.3$--$1.9$ \citep{hartmann1998,muzerolle2003,muzerolle2005,calvet2004,mohanty2005,natta2006,garcialopez2006,gatti2008,biazzo2012,fang2009,fang2013,ercolano2013}. In the Orion Nebula Cluster (ONC) \citet{manara2012} found a typical trend for the entire stellar sample ($\eta\sim1.4$), from a uniquely large sample of precise $\dot M_{\rm acc}$ estimates (over 700 sources). These measurements were derived combining Hubble Space Telescope photometry of stellar flux excesses with the available spectroscopically determined stellar photospheric parameters in the ONC from \citet{dario2012}. Despite the typical value of $\eta$ for the overall population, \citet{manara2012} also found that stars of about a solar mass showed a systematic slower decay of accretion, with $\eta\sim0.5$. Similar results for intermediate  masses ($M\gtrsim0.5$~M$_\odot$) were obtained in several  detection-limited photometric studies of $\dot M_{\rm acc}$ in the  Magellanic Clouds  \citep[e.g.,][]{demarchi2010,demarchi2011a,demarchi2011b,spezzi2012}  as well as in some Galactic star forming regions  \citep[][]{beccari2010,rigliaco2011,demarchi2012}. These findings suggest that solar-type stars may sustain mass accretion from long-lived gaseous  disks on timescales $>10$~Myr.

The measurement of accretion rates, as well as that of the stellar parameters of young stars, are affected by large uncertainties. As we will show, such uncertainties, together with observational obstacles such as incompleteness due to detection limits, influence the measured dependence of $\dot M_{\rm acc}$ on stellar properties.
In this paper we focus on the biases in the measured correlations between $\dot M_{\rm acc}$ and stellar age, and the estimated accretion timescales. In what follows we will characterize this bias by comparing the true value of $\eta$ in equation~\ref{equation:mdot_propto}, which we refer to as $\eta_{\rm true}$, with that obtained from fitting the $\dot M_{\rm acc}$ versus $\log t$ relationship to observational or simulated data, which we refer to as $\eta_{\rm meas}$.
In Section \ref{section:the_problem} we introduce the major underlying problems with measuring $\dot M_{\rm acc}$, stellar parameters, and thus $\eta$; in Section \ref{section:simulations} we set out our modelling approach and the underlying assumptions. In Section \ref{section:results} we present the results of simulations to study the systematic biases in $\eta$ introduced by a combination of uncertainties and selection effects. A more detailed individual treatment of each of these effects is presented in Appendices \ref{appendix:simulations_simplespreads} and \ref{section:selection_effects}. Finally we discuss implications and caveats of our findings (Section \ref{section:discussion}) and summarize (Section \ref{section:summary}) our conclusions.

\section{The problem}
\label{section:the_problem}
\subsection{Stellar parameters of young stars are uncertain}
\label{section:the_problem_age}
The stellar parameters of young stars usually cannot be assigned with good accuracy, due to both observational uncertainties and the physical nature of PMS stars \citep[see the review by][]{preibish2012}. Young clusters are often partially embedded in their parental clouds, causing differential extinction between their members; $T_{\rm eff}$ and $A_V$ must be estimated for individual stars based on either spectroscopic or photometric techniques. PMS stars are also known to exhibit substantial variability \citep{herbst2002}, due to rotation of heavily spotted photospheres and non steady accretion. Moreover, the continuum emission produced by accretion is often not properly characterized \citep{dario2010a,manara2013} and it may present peculiar spectral energy distributions \citep{edwards2006,fischer2011}. Unresolved stellar multiplicity also biases the observed luminosities and inferred ages \citep{dario2010b,naylor2009}.

Furthermore, the history of protostellar accretion is thought to
influence the subsequent stellar structure for the first few Myrs of PMS
evolution \citep{baraffe2009,baraffe2012}. If different stars of a
population undergo a diversity of protostellar buildup histories during
the embedded phase
(intensity of accretion bursts, different initial conditions), this can
result in a significant luminosity spread even for coeval stars.
The effect is most prominent for ``cold''
protostellar accretion - when little or no energy from the accretion
flow is stored in the stellar interiors, and may result in
systematically smaller radii and lower luminosities for stars of $\sim 1\,M_{\odot}$
even several Myr after they emerge as T-Tauri stars \citep{hosokawa2011}.
Stellar radii may also expand in response to ``hot'' protostellar accretion,
but the contraction timescales will then be so short ($\ll 1$~Myr) that such objects
are unlikely to contribute significantly to the luminosity spread seen
in a population of PMS objects.

Whereas both stellar masses and ages are uncertain due to these issues, ages are far more problematic: first, protostellar accretion produces spreads mostly in $\log L_*$, which correlates with $t$ and biases $M_*$ only marginally; second, the typical range of masses observed in PMS clusters is large ($>1$~dex), and much larger than the uncertainties on this quantity.

All these effects broaden the measured age distribution in young clusters and lead to overestimates of the real intrinsic age spread. The actual extent of this effect is partially unknown, but some studies carried out in the ONC, have tried to constrain it. The population of the ONC, when stars are carefully positioned in the HRD using spectroscopic spectral types complemented with photometry to disentangle dust extinction for individual sources, shows an isochronal spread in log age of $\sigma \log t \simeq0.4$~dex  around a mean, model dependent, age of $\langle t\rangle\sim 2$~Myr \citep{dario2010a,dario2012}. Independent studies based on the analysis of uncertainties \citep{reggiani2011} and projected stellar radii \citep{jeffries2007} indicate that most of the luminosity spread must be attributed to a spread in stellar radii, and the observational uncertainties have a marginal role, accounting for an apparent age spread of only $\sigma \log t_{\rm errors}\sim0.16$. Nevertheless, \citet{jeffries2011} points out that not all the spread in radii can be attributed to an intrinsic age spread, since an expected correlation between isochronal ages and disk fractions is not observed. Through statistical simulations, they estimate an upper limit on the real age spread in the ONC $\sigma\log t\leq0.2$~dex. This result, combined with evidence for a true spread in radii and the small impact of observational uncertainties, suggests that a ``non negligible'' luminosity spread may be induced by protostellar accretion histories. Based only on spatial and kinematic considerations, a real age spread must exist at some level \citep[e.g.,][]{tan2006} and extreme perturbations in the HRD due to protostellar accretion histories are not supported by measurements of lithium depletion \citep{sergison2013}.

\subsection{Measurements of $\dot M_{\rm acc}$ are uncertain}
\label{section:the_problem_mdot}
The gravitational energy released by infalling material produces a characteristic flux excess over that expected from the stellar photosphere \citep{calvet-gullbring1998,gullbring1998}. This comprises a strong UV excess produced by the shock at the photosphere, wide optically thin recombination lines from the infalling column, and optical continuum from the heated photosphere, often referred to as veiling. Practically, the bolometric accretion luminosity $L_{\rm acc}$ is estimated by measuring the intensity of a portion of the excess emission, usually hydrogen ricombination lines -- especially H$\alpha$ in the optical-- or the Balmer jump excess in the $U$-band, from either spectroscopy \citep[][]{gullbring1997,valenti1993,herczeg-hillenbrand2008,gatti2008,biazzo2012,fang2009,fang2013,rigliaco2012,manara2013,ingleby2013,alcala2013} or photometry \citep{robberto2004,romaniello2004,demarchi2010,manara2012}.

Practically, the measurement $L_{\rm acc}$ is some cases imprecise. Separating $L_{\rm acc}$ from the total stellar flux requires a good knowledge of the photospheric flux, but this is often imperfect (see Section \ref{section:the_problem_age}). In reality, the system of star, disk and accretion have a more complex interplay between the radiative and hydrodynamical aspects \citep[e.g.,][]{tannirkulam2008,mayne-harries2010}. Also, the bolometric $L_{\rm acc}$ is extrapolated from portions of its continuum or line emission based on empirical relations \citep{herczeg-hillenbrand2008,dahm2008,rigliaco2012,alcala2013}, which are affected by a significant dispersion, and in some cases the spectrum of accretion excess can present large anomalies \citep{edwards2006,fischer2011}.

Even when the measurements of $L_{\rm acc}$ are precise, their conversion into mass accretion rates requires additional assumptions. Since $L_{\rm acc}$ originates from the release of gravitational energy,
\begin{equation}
L_{\rm acc}=G\frac{\dot M_{\rm acc} M_*}{R_*}\bigg(1-\frac{R_*}{R_{in}}\bigg)
\end{equation}
 \noindent where $R_*$, $L_*$ and $M_*$ are the stellar radius, luminosity and mass, and $R_in$ the inner disk radius. Thus, since it is commonly assumed $R_{in}\sim5R_*$:
\begin{align}
\dot{M}_{\rm acc}&=\frac{L_{\rm acc} R_*}{0.8GM_*} \label{equation:mdot1}\\
&=\bigg(\frac{ L_{\rm acc}}{L_*}\bigg)\frac{L_*R_*}{0.8GM_*}\ . \label{equation:mdot2}
\end{align}

In reality $R_{in}$ is probably
closer to the co-rotation radius \citep{shu1994,koenigl1991}, which
varies star by star as the typical PMS rotation periods $\tau$ range
from less than a day to $\sim10$ days. This produces uncertainties of a
factor 2 or more in estimates of $\dot{M}_{\rm acc}$.

Fortunately, the uncertainties in estimates of $\dot M_{\rm acc}$
should be a minor cause of bias in $\eta$ compared to uncertainties in
the stellar parameters (Section \ref{section:the_problem_age} and
\ref{section:the_problem_correlated_uncertainties}). First, estimates
of $\eta$ are generally based on a large number of sources. For
example, the work of \citet{manara2012} determined $\dot M_{\rm acc}$
in the ONC for over 700 sources, using HST photometry to measure both
H$\alpha$ and $U$-band excess. The large number statistics mitigate
against the inevitable presence of peculiar sources departing from
the assumptions of Equation \ref{equation:mdot1}, (although this does not preclude systematic problems with this equation). Second, in the majority of the studies cited in
Section~\ref{section:intro}, $\dot M_{\rm acc}$, and hence $\eta$, was estimated for sources whose stellar parameters have been assigned from spectroscopy; thus large anti-correlations between $L_*$ and $L_{\rm acc}$ from a lack of knowledge on what fraction of the photospheric flux is to be attributed to accretion should not be present.

\subsection{Correlated uncertainties between  $\dot M_{\rm acc}$ and stellar parameters}
\label{section:the_problem_correlated_uncertainties}
\begin{figure*}
\includegraphics[width=1.6\columnwidth]{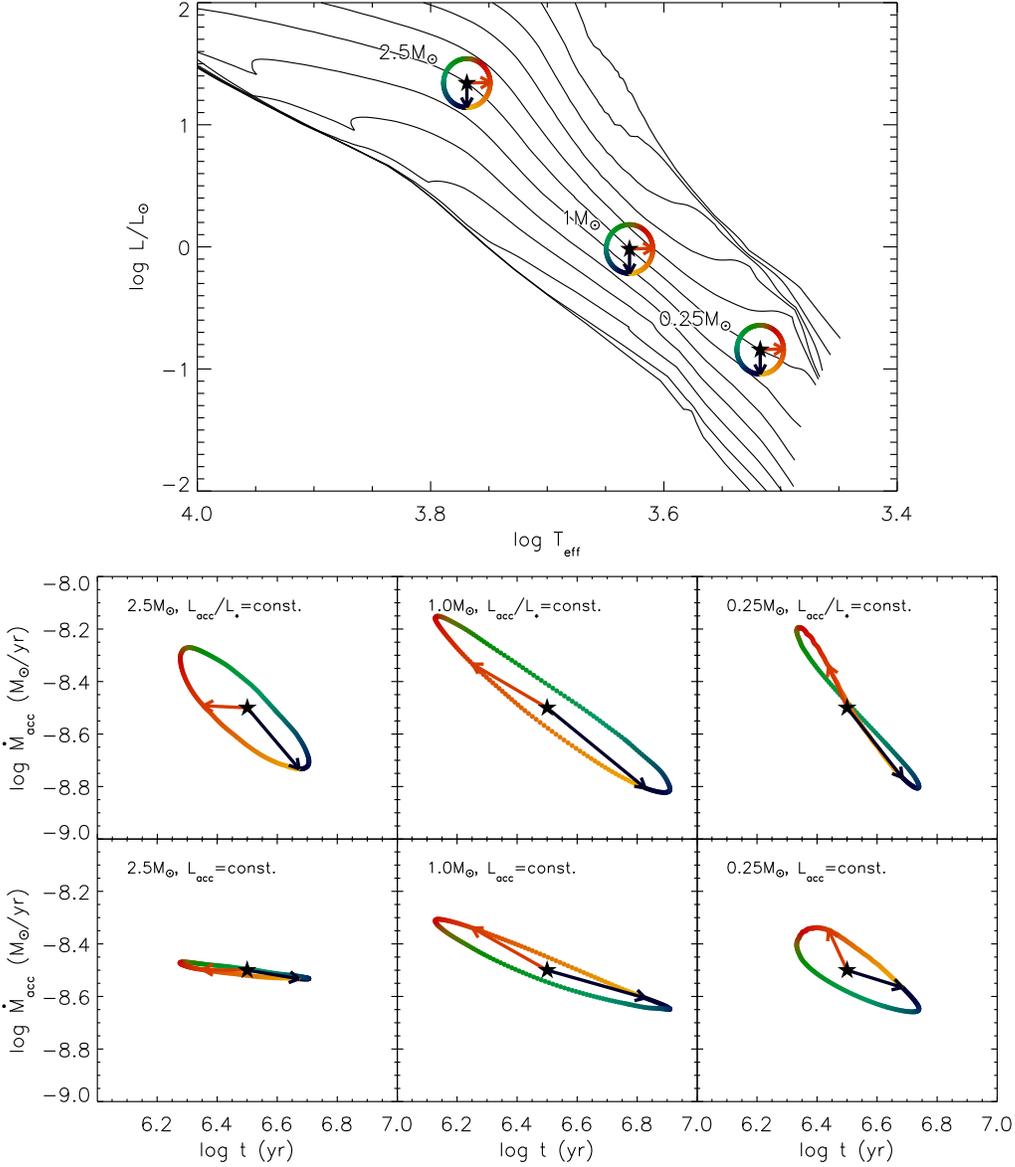}
\caption{Examples of the correlated propagation of uncertainties from   the HRD to the $\dot M_{\rm acc}$ versus age plane for three test   stars with age $t=3$~Myr, masses of 0.25, 1 and 2.5~$M_\odot$, and   $\dot M_{\rm acc}=10^{-8.5}$. When we displace the HRD position   (\emph{top panel}) of a star to a circle around the true $T_{\rm     eff}$ and $L_*$, the estimated $\dot M_{\rm acc}$ and age are   displaced as in the bottom panels. The middle three panels assume   that $L_{\rm acc}$ is measured accurately (Case A), the bottom three panels   assume that $L_{\rm acc}/L_{\rm bol}$ is measured accurately (Case B).   Points of same colors and arrows   correspond between HRD and the $\log \dot M_{\rm acc}$--$\log t$   panels. \label{fig:hrd_example}}
\end{figure*}
If the errors in $\dot M_{\rm acc}$ are symmetric and independent of the stellar parameters, which is largely the case for several of the issues described above, only a vertical scatter is introduced in the measured $\log \dot M_{\rm acc}$--$t$ relation, with no effect on $\eta$.
However, equation \ref{equation:mdot1} shows that $\dot M_{\rm acc}$ is strongly dependent on the assigned stellar parameters $R_*$,$L_*$ and $M_*$. Since, at a given $T_{\rm eff}$, $L_*\propto R_*^2$, Equation \ref{equation:mdot2} implies that  $\dot M_{\rm acc}\propto R_*^3$. Moreover, for low-mass PMS stars stellar evolutionary models predict that $M_*\sim T_{\rm eff}^6$, and thus: \begin{equation}
\dot M_{\rm acc} \propto \bigg(\frac{ L_{\rm acc}}{L_*}\bigg)\frac{R_*^3 T_{\rm eff}^4}{M_*} \sim \bigg(\frac{ L_{\rm acc}}{L_*}\bigg)L_*^{\frac{3}{2}} T_{\rm eff}^{-2}
\end{equation}
During the PMS contraction towards the main sequence, $L_*$ decreases roughly as $L_*\propto t^{-2/3}$, while $M_*$ depends on both $L_*$ and $T_{\rm eff}$. Therefore, if $L_*$ or $T_{\rm eff}$ are uncertain, the consequent errors in age and mass will correlate with those in $\dot M_{\rm acc}$.

The bias in $\dot M_{\rm acc}$ for a given star depends on the actual interplay between uncertainties in stellar parameters and in $L_{\rm acc}$, which is in turn dictated by the physical or observational effects altering the estimate of these quantities, and can vary from star to star. We can however treat this problem in a simplified way and describe 3 limiting cases:
\begin{itemize}
\item \textbf{\emph{Type A}}: The ratio $L_{\rm acc}/L_*$ is not affected by the uncertainty in the stellar parameters. This assumption is adequate for a number of scenarios where the presence of an error in the estimated stellar luminosity affects $L_{\rm acc}$ in the same way; e.g.: 1) wrong or uncertain distance to a particular source; 2) when a fraction of the stellar light is blocked, by, e.g, a circumstellar disk seen nearly edge-on; 3) when $T_{\rm eff}$ is correctly estimated, but the extinction in the band used to obtain $L_*$ is incorrect because of a peculiar reddening law; 4) in the case of unresolved binarity and both components are (equally) accreting.
    Also dust extinction does not significantly affect the ratio $L_{\rm acc}/L_*$, especially if $\dot M_{\rm acc}$ is measured from the $H\alpha$ excess \citep{demarchi2010}.

    From Equation \ref{equation:mdot2} we then have:
    \begin{equation}
    \label{equation:typeA}
    \dot M_{\rm acc}^{\rm meas}=\dot M_{\rm acc} \cdot \frac{R_{\rm meas}}{R_*}\frac{L_{\rm meas}}{L_*}\frac{M_*}{M_{\rm meas}}
    \end{equation}
\item \textbf{\emph{Type B}}: the accretion luminosity $L_{\rm acc}$ is
  correctly measured, even if the $L_*$ is wrong. This is the case for
  errors in the stellar luminosity such as: 1) rotational variability
  due to dark spots; 2) incorrect accounting for optical continuum
  excess (veiling),
%or peculiar one; **not sure what you mean here? RDJ**
3) unresolved multiplicity with only one companion responsible for the measured $L_{\rm acc}$. From Equation \ref{equation:mdot1} we have:
    \begin{equation}
    \label{equation:typeB}
    \dot M_{\rm acc}^{\rm meas}=\dot M_{\rm acc} \cdot \frac{R_{\rm meas}}{R_*}\frac{M_*}{M_{\rm meas}}
    \end{equation}
\item \textbf{\emph{Type C}}: a variety of protostellar accretion histories \citep{baraffe2012} produces a range of possible $R_*$ for the same $M_*$ and age, leading to an observed luminosity spread. Here we assume that $T_{\rm eff}$ is unperturbed. Unlike types A and B, this is not an error in the position of the star in the HRD, as $T_{\rm eff}$, $L_*$, $R_*$ are correctly determined; rather, $t_*$ and $M_*$ are wrong, since they are assigned using a unique, incorrect grid of evolutionary models with a single birthline. Thus, the biased estimate of mass accretion rates will be:
    \begin{equation}
     \label{equation:typeC}
    \dot M_{\rm acc}^{\rm meas}=\dot M_{\rm acc} \cdot \frac{M_*}{M_{\rm meas}}
    \end{equation}
    \noindent where $M_{\rm meas}$ is the mass assigned to the new position in the HRD using non-accreting isochrones.
\end{itemize}

In reality, when a real stellar population is considered, the overall effect of uncertainties will be due to a combination of biases of types A,  B and C.
In principle, following the definition of the type A and B biases, an additional type of uncertainty is present; this is when $L_*$ is correct but $L_{\rm acc}$ is not. This scenario, which encompasses most of the difficulties with the measurement of $L_{\rm acc}$ we discussed in Section \ref{section:the_problem_mdot}, is however irrelevant for our analysis: if the photospheric parameters are not perturbed, the ages are unaffected, and the result is a vertical (i.e. uncorrelated) spread in the $\log \dot M_{\rm acc}$ versus $t$ plane, which does not alter $\eta_{\rm meas}$. This will occur when $L_{\rm acc}$ uncertainties are uncorrelated with the stellar parameters; for instance, where the ``non universality'' -- or contamination -- of the accretion spectrum \citep{edwards2006,fischer2011,manara2013} or the temporal variability of $\dot M_{\rm acc}$ alter the estimated $L_{\rm acc}$ \citep{bouvier2003,alencar2005,kurosawa2008,nguyen2009,costigan2012}.

Figure \ref{fig:hrd_example} illustrates an example of how uncertain stellar parameters bias the measured $\dot M_{\rm acc}$--$t$ correlation. We consider 3 stars on a 3~Myr \citet{siess2000} isochrone, with masses $M_*=0.25$, 1 and 2.5~M$_\odot$, and assign them an arbitrary value of $\log \dot M_{\rm acc}=-8.5$ in units of $M_{\odot} $yr$^{-1}$. We then perturb $T_{\rm eff}$ and $\log L_*$, shifting their values along circles in the HRD, mimicking the effect of an uncertainty of 0.2~dex in $\log L_*$ and 0.02~dex in $\log T_{\rm eff}$. For each of these points, we re-assign $M_*$ and $t$ by isochrone interpolation, and derive the correction to $\dot M_{\rm acc}$ due to the resulting change in the stellar parameters $R_*,L_*,M_*$ from Equation \ref{equation:mdot2}, assuming a bias of either type A or B. The result shows the highly correlated shifts in $\log t$ and $\log \dot M_{\rm acc}$ produced by errors in $\log T_{\rm eff}$ and $\log L_*$, with a slope (on logarithmic axes) ranging from 0 to $-1$.

The examples shown in Figure \ref{fig:hrd_example} suggest that when an entire stellar population is considered, the data in the $\log \dot M_{\rm acc}$--$\log t$ plane are shifted by uncertainties along some preferential direction, affecting the measured slope $\eta_{\rm meas}$ and introducing an apparent correlation between these two quantities even in the limiting case of a coeval population.
Since, as we discussed earlier, uncertainties in the isochronal ages in young clusters are large, perhaps accounting for half or more of the width of the apparent age distribution, this bias in $\eta$ may be significant. Also, given that ages are generally placed on the $x$-axis to determine $\eta$ from the $\log \dot M_{\rm acc}$--$\log t$ plane, regression dilution (\citealt{frost2000}) becomes important. This is a general phenomenon affecting linear regression and a consequence of it being a non symmetric approach in the two variables. Statistical variability, measurement error or random noise in the $y$ variable cause uncertainty in the estimated slope, but not bias; on the other hand any noise, errors or scatter in the independent variable $x$ does bias the estimated slope, and the greater the variance in the $x$ measurement (our stellar age), the closer the estimated slope (our $\eta_{\rm meas}$) approaches zero instead of the true value. This bias could in principle be removed, but only if error bars in both axes are well known and accounted for when fitting the data. This is not generally performed for stellar ages.

In the analysis of this paper (see Section \ref{section:results}) we will assume that the bias in $\dot M_{\rm acc}$ affecting $\eta$ originates from the errors in the stellar parameters, and that any uncertainty from the $L_{\rm acc}$ estimate is independent of the stellar parameters. However, in some cases the interplay between the uncertainties is more complex and may produce additional biases in $\eta$. When the data are insufficient to properly separate the photospheric and accretion contribution to the total flux -- e.g., when results are based solely on broad-band photometry, and for intrinsically faint sources like brown dwarfs -- an underestimate of $L_{\rm acc}$ leads to an overestimate of the stellar $T_{\rm eff}$ and $L_*$, with a consequent simultaneous error in both age and $M_{\rm acc}$ \citep{mayne-harries2010}. Additionally, as stellar rotation periods in young stars vary with isochronal age \citep{littlefair2011}, some correlation between age and inner (co-rotation) disk radius may be present, introducing an age-dependent variation of the term $1-R_{*}/R_{in}\sim0.8$ in Equation \ref{equation:mdot1}. Similarly, stellar rotation in young clusters often appears slower for disk-bearing stars \citep[e.g.][]{rebull2006,biazzo2009}, and the presence of disks is both directly related to the accretion process, and a contributor to the assignment of inaccurate stellar parameters \citep{guarcello2010,kraus-hillenbrand2009}. All of these factors are yet to be sufficiently characterized or understood well enough to study their possible biasing impact on the measured time decay of $\dot M_{\rm acc}$. Because of this, and as in this work we aim to look for systematic biases in the published values of $\eta_{\rm meas}$, which usually assume Equations \ref{equation:mdot_propto} and \ref{equation:mdot1}, we will focus on the bias produced by uncertain stellar parameters, and leave additional comments addressing the possiblity of other effects to the discussion in Section \ref{section:discussion}.

\subsection{Selection effects due to incomplete samples}
Selection effects due to accretion detection thresholds may also bias $\eta_{\rm meas}$. Small $\dot M_{\rm acc}$ is detected more easily in faint stars than in bright stars, since the contrast, $L_{\rm acc}/L_*$, is larger.
This issue has been pointed out as critical, especially for measuring the mass dependence of $\dot M_{\rm acc}$ through the parameter $b$ in equation~\ref{equation:mdot_propto} \citep{clarke2006,mayne-harries2010}. The measured $\dot M_{\rm acc}$--$t$ correlation is also affected since $L_*$ decreases with apparent age. Alternatively, if the stellar sample is incomplete at faint luminosities, older stars with small $\dot M_{\rm acc}$ could be systematically missed.

A flat detection limit in stellar luminosity (or magnitude) and a uniform threshold in the quantity used to determine accretion excess (e.g., H$\alpha$ equivalent width, or $U-$ band excess in magnitudes) will result in an inclined detection threshold in the $\dot M_{\rm   acc}$--$t$ plane, skewing the relationship between these quantities deduced from samples with detected accretion. This type of selection effect is dominant in studies of distant clusters (e.g., the Magellanic Clouds, or extincted Galactic clusters), where sources cannot be detected in the very low-mass range, and in older $\gtrsim10$~Myr clusters, where the fraction of sources with detectable mass accretion is minimal.

\section{The method}
\label{section:simulations}
\begin{figure*}
\includegraphics[width=\columnwidth]{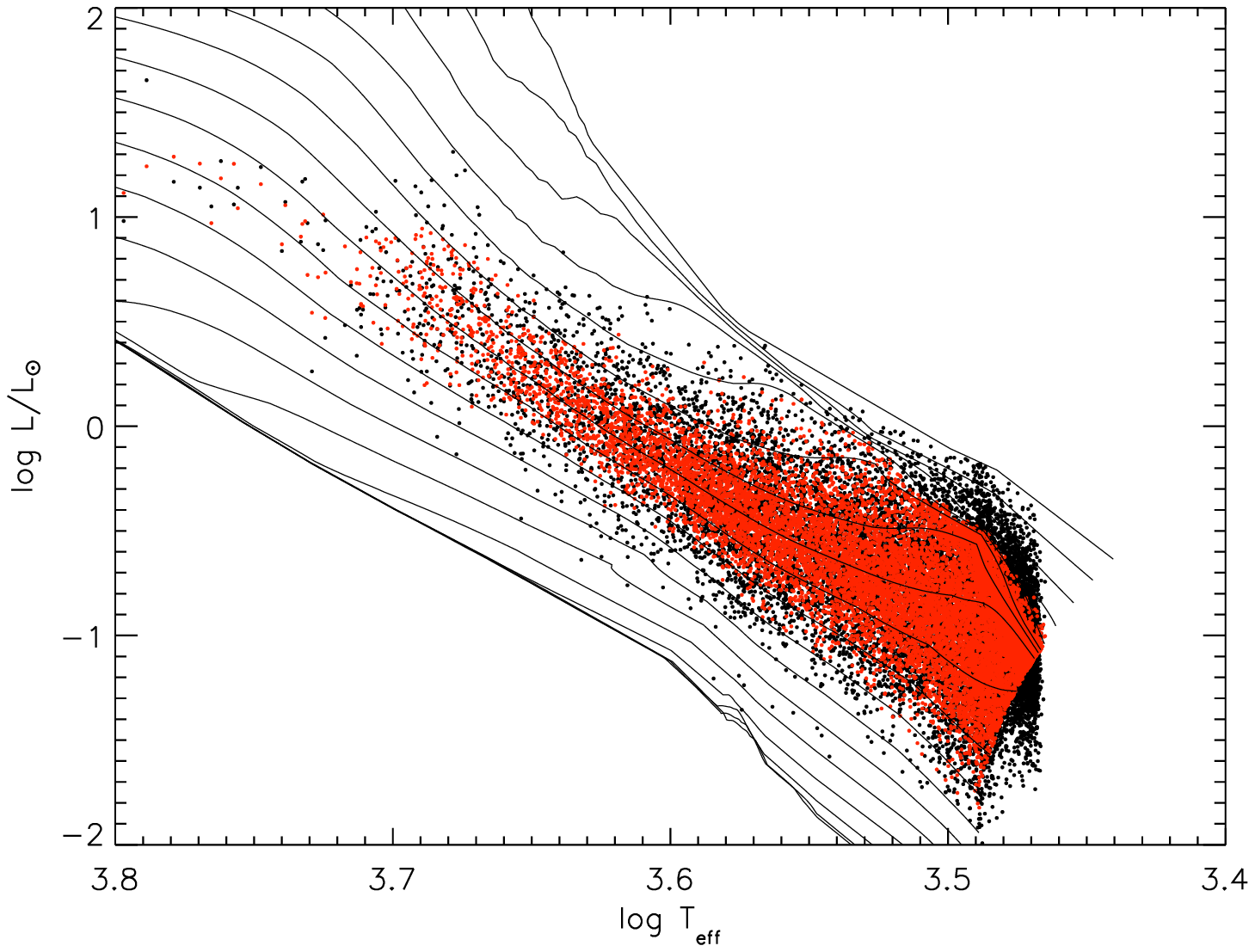}
\includegraphics[width=\columnwidth]{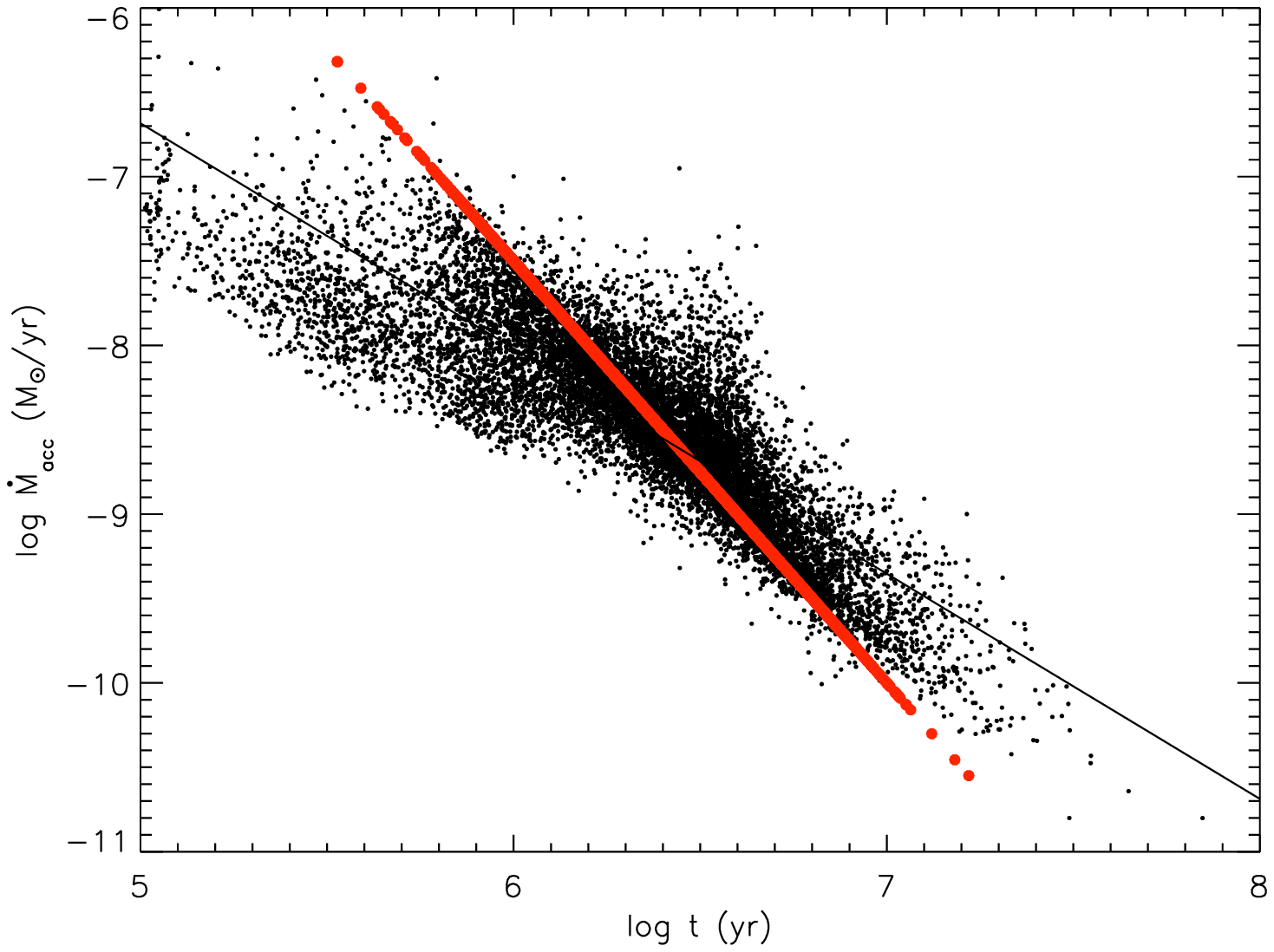}
\caption{{\emph Left:} An example of a HRD simulation. The red points   represent the input population, which has an intrinsic age spread of   $\sigma\log t=0.2$~dex. The black points indicate are the same stars   after the application of a further random gaussian luminosity spread   (see text)   to produce the observed HRD, which has $\sigma\log t\simeq0.4$~dex. {\emph Right:} The effect in the $\dot M_{\rm acc}$--$t$ plane. The red dots show the input values ($\dot M_{\rm acc}\propto t^{-2.5}M_*^0$); the black dots are the recovered values after both $t$ and $\dot M_{\rm   acc}$ are calculated from the observed HRD position. The solid line is the best fit ($\dot M_{\rm acc}^{\rm meas}\propto t_{\rm meas}^{1.34})$. \label{fig:simplespread_mdot_age}}
\end{figure*}

We perform simulations of synthetic populations in the HRD to
characterize the biases in the measured exponent $\eta_{\rm meas}$, due
to inaccurate assignment of stellar parameters and selection
effects. The general method is straightforward: we generate a large set
of PMS stars assuming an age distribution; we assign values of $\dot
M_{\rm acc}$ to each star according to an assumed value of $\eta_{\rm
  true}$ and $b$. Then we displace the sources in the HRD mimicking the
effect of different types of uncertainties. For each source, its
perturbed position in the HRD results in the assignment of a new
measured age from isochrone interpolation, as well as a new measured
$\dot M_{\rm acc}$ using the perturbed stellar parameters, $R_*$,
$L_*$, $M_*$, in Equation \ref{equation:mdot2}). Lastly, $\eta_{\rm meas}$ is found by linear regression and compared to the assumed $\eta_{\rm true}$.

\subsection{General assumptions}
\label{section:simulations_general_assumptions}

Throughout our simulations, we assume a cluster population with a mean age of 2.5~Myr, similar to that of the ONC when \citet{siess2000} isochrones are used to assign ages \citep{dario2010b}. In fact the absolute age of Orion, as that of any young cluster, is uncertain by a factor of several and model dependent. However, the ONC has a relative age comparable, within a factor of 2, to the majority of star forming regions where accretion timescales are studied. Our assumed mean cluster age of 2.5~Myr is purely dictated by our intention to generate an observed HRD comparable with the ONC, since we will adopt \citet{siess2000} models, but its absolute value has little to do with the phenomena we aim to study.
Since PMS isochrones in the HRD are almost parallel and spaced by nearly identical offsets in $\log t$, changing the mean age for a population produces mostly a horizontal shift in the $\dot M_{\rm acc}$--$t$ plane with little effect on $\eta_{\rm meas}$. In Section \ref{appendix:evolmodels} we demonstrate this, and show that the behavior of the bias in $\eta$ remains nearly identical if another set of evolutionary models with a large difference in predicted absolute age scale is adopted instead of the \citet{siess2000} isochrones.

Although $\eta_{\rm meas}$ is insensitive to changes in the absolute age scale, the measured accretion timescale $\tau_{\rm acc}\simeq \langle t\rangle/\eta_{\rm meas}$ is proportional to the absolute age of the system. This is a general problem for the study of young cluster evolution, with consequences extending far beyond the scope of this paper \citep{naylor2009}.

In several parts of this work, e.g., when studying the effect of
luminosity spreads alone (Appendix
\ref{section:simulations-simplespread}), and when multiple combined
uncertainties are considered (Section \ref{section:combined effects}),
the apparent broadening of the HRD is tuned to reproduce the apparent
age spread $\sigma\log t_{\rm meas}=0.4$ observed in the ONC, as well
as its observed mean age, and the value of $\eta_{\rm meas}$ in the
region. None of these quantities are peculiar to the ONC, and other
Galactic star forming regions share similar values,
so our treatment and results should be of general applicability.

We assume that the observed spreads in apparent age are produced by a combination of a real age spread and spreads due to observational uncertainties that are formed by displacing sources in the HRD along the $\log L_*$ axis, the $\log T_{\rm eff}$ axis, and along both axes due to uncertain $A_V$. We let the real spread in log age range between $\sigma\log t=0.1$~dex and 0.3~dex, tuning the observational uncertainties to produce the observed apparent log age spread of 0.4~dex. This leaves open the possibilities that the additional spread in apparent age, beyond that attributable to the {\it known} observational uncertainties \citep[e.g.,][]{reggiani2011}, is due to (a) physical effects that lead to different HRD positions for stars of the same age (e.g. the protostellar accretion history) or (b) further unrecognised sources of uncertainty in $L_*$ and $T_{\rm eff}$ \citep[e.g.,][]{manara2013}.

Lastly, we initially assume that $\dot M_{\rm acc}$ depends solely on
the stellar mass and age through
equation~\ref{equation:mdot_propto}. As anticipated in Section
\ref{section:the_problem_correlated_uncertainties}, other physical and
observational effects may add further complications if the error in
deriving $L_{\rm acc}$ has some correlation with the assigned stellar
parameters. However, we deliberately base our work on the ONC to
mitigate such scenarios. The ONC,
is perhaps the most extensively studied young cluster, where stellar
parameters and accretion luminosities have been assigned with the
highest achievable precision and completeness, and are subject to the
minimum of observational bias.
In Section \ref{section:discussion} we discuss the implications of
relaxing this assumption.

\subsection{The simulations}
\label{section:simulation_howtheywork}
We generate 50,000 random stars with ages drawn from the assumed age distribution and masses drawn from the \citet{kroupa2001} initial mass function. We then assign their $T_{\rm eff}$, $L_*$, $R_*$ by interpolation of the \citet{siess2000} evolutionary models. The assigned $\dot M_{\rm acc}$, following a chosen $\eta_{\rm true}$ (generally between 0 and 4) and $b$ (generally either 0 and 2), requires an arbitrary normalization: we impose $\log \dot M_{\rm acc}=-8.5$ for $M_*=0.5$~M$_\odot$ and $t=2.5$~Myr. A different normalization would only produce a rigid vertical shift in the $\log \dot M_{\rm acc}$--$\log t$ plane, with no effect on $\eta_{\rm meas}$

We then randomly displace the positions of these sources in the HRD to reproduce the effect of a given observational uncertainty (e.g., an assumed apparent luminosity spread, or a differential extinction), or physical mechanism (e.g., a distribution of $R_*$ from protostellar accretion history). In any case, the resulting (broader) HRD represents the observed one, thus, for each star, the measured stellar parameters ($M_{\rm meas}$, $t_{\rm meas}$, $R_{\rm meas}$), altered from the true ones, are assigned by interpolation of the Siess evolutionary models in the perturbed HRD. As for the measured value of $\dot M_{\rm acc}$ for each source (see Equation \ref{equation:mdot1}), the way it changes compared to the true value assigned to each star depends on the nature of the particular uncertainties affecting the stellar parameters, and is computed from Equations \ref{equation:typeA}, \ref{equation:typeB} and \ref{equation:typeC}.

Figure \ref{fig:simplespread_mdot_age} illustrates an example of our simulations. This assumes to begin with a real age spread $\sigma \log t=0.2$~dex (the upper limit in the real age spread for the ONC from \citealt{jeffries2011}), $\eta_{\rm true}=2.5$, $b=0$; a gaussian random luminosity spread with $\sigma \log L_*=0.2$~dex is added, and the uncertainties in $\dot M_{\rm acc}$ are of type A. The resulting age distribution, from isochrone interpolation after applying the luminosity spread, has a width $\sigma \log t_{\rm meas} \simeq 0.4$~dex, similar to that measured in the ONC. The best linear fit to the population in the $\log \dot M_{\rm acc}$--$\log t$ has a slope $\eta_{\rm meas}=1.34$, considerably shallower than $\eta_{\rm true}$.

\section{Results}
\label{section:results}
\subsection{Individual uncertainties and selection effects}
We start by separately considering the bias in $\eta$ from individual types of uncertainty in the HRD. These are respectively, uncertainties in luminosity (including those due to variations in protostellar accretion history), uncertain $T_{\rm eff}$ or reddening, and systematic mass-age correlations introduced by evolutionary models. The full description of our methods, assumptions, and findings are in Appendix \ref{appendix:simulations_simplespreads}; here we just summarize.
\begin{itemize}
\item Luminosity spreads in the HRD strongly affect the measured
  correlation between $\dot M_{\rm acc}$ and age, resulting in
  $\eta_{\rm meas} < \eta_{\rm true}$.
\item If we assume an intrinsic age spread $\sigma\log t=0.2$ (half of
  the total apparent one), then $\eta_{\rm true} \simeq
  3 \eta_{\rm meas}$. If the observed luminosity dispersion has a smaller contribution
  from a real age spread, then this bias becomes larger, and
  vice-versa.
\item If the luminosity spreads originate from changes in radius (at a
  given age) due to varying protostellar accretion histories rather than
  observational uncertainties, then $\eta_{\rm true}/\eta_{\rm meas}$
  becomes even larger.
\item Uncertainties in $T_{\rm eff}$ and extinction $A_V$ also result in
 $\eta_{\rm meas} < \eta_{\rm true}$; however, for realistic values,
  the effects are more modest than those due to luminosity uncertainties.
\item Changing family of stellar evolutionary models has very little effect on the bias in $\eta$, even when large systematic differences in the absolute age scale are present. However, a large tilt in the isochrones, as needed to correct them for the typically observed mass-age correlation in their prediction, further increases the bias in the same direction as produced by uncertainties.
\end{itemize}

In Appendix \ref{section:selection_effects} we study in detail the bias
in the $\dot M_{\rm acc}$--$t$ correlation introduced by  source
detection incompleteness and a threshold for the measurement of $\dot
M_{\rm acc}$. These selection effects generate apparent correlations in
the mass-age and
$\log \dot M_{\rm acc}$--$\log t$
planes for those sources which are detected and for which accretion can be measured. Our simulations show that as a result, $\eta_{\rm meas}$ is displaced towards a value of 1. Since $\eta_{\rm meas}$ in young clusters is commonly in the range $1<\eta_{\rm meas}<2$, a larger $\eta_{\rm true}$ is required to match observations when these selection effects are present.

\subsection{Combined effects}
\label{section:combined effects}

\begin{figure*} \includegraphics[width=1.6\columnwidth]{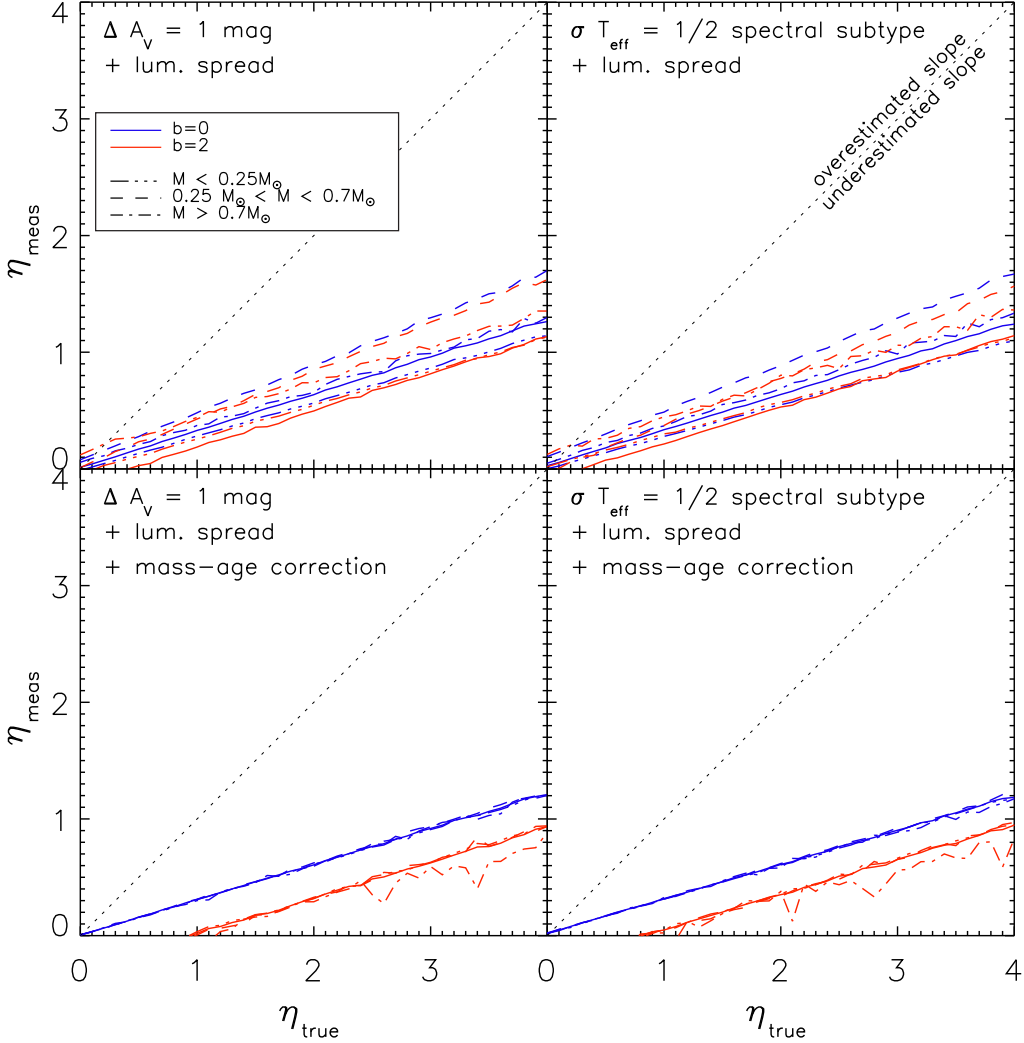} \caption{Measured $\eta_{\rm meas}$ versus $\eta_{\rm true}$, for a   2.5~Myr population with an age spread $\sigma \log t_{\rm true}=0.2$~dex and   additional uncertainties producing a total apparent age spread of   $\sigma \log t_{\rm meas}=0.4$~dex, from the combined effects of a luminosity error and 1~mag of differential $A_V$ (\emph{left panels}) or $T_{\rm eff}$ uncertainty of half spectral subtype (\emph{right panels}), and including the systematic offset in luminosity to correct for the apparent mass-age correlation introduced by the use of non-accreting isochrones (\emph{bottom panels}). \label{fig:tautau_lumi_teff_underl}} \end{figure*}

Following the analyses in Appendices \ref{appendix:simulations_simplespreads} and \ref{section:selection_effects} where individual biasing effects on $\eta$ are studied, we now analyse the overall effect of realistic simultaneous combinations of different sources of uncertainty.

We start by considering the simultaneous effect on $\eta$ of both uncertainties in $T_{\rm eff}$ or $A_V$ and luminosity spreads, but ignore incompleteness due to selection effects. We assume a real age spread $\sigma \log t=0.2$~dex around a mean age of 2.5~Myr and perform simulations of synthetic populations as described in Section \ref{section:simulation_howtheywork}, proceeding in two steps. First we randomly displace the stellar parameters by adding an uncertainty in $T_{\rm eff}$ of half a spectral subtype, or, separately, an uncertainty in $A_V$ with a flat distribution $0<\Delta A_V<1$~mag (which can also be thought of as neglecting to take account of differential extinction ranging between 0 and 1 mag). In Appendices \ref{section:simulations-teff} and \ref{section:simulations-teffav} these effects are treated in isolation, but here we assume that these uncertainties will inevitably lead to correlated uncertainties in the luminosity estimate. In other words, an error in $A_V$ leads to an error in $\log L_*$, and an error in $T_{\rm eff}$ will result in an incorrectly estimated reddening and extinction, and hence an error in $\log L_*$. Both of these shift the simulated stars in the same direction in the HRD but the extent of the shifts depends on the respective $A_V$ or $T_{\rm eff}$ errors.

In a second step, we add a further luminosity uncertainty, by applying
a gaussian shift in $\log L_*$ for each star, with a standard deviation
tailored to obtain an apparent age spread $\sigma \log t=0.4$~dex. We
consider a combination of uncertainties of types A, B and C for this
additional luminosity spread, taking the ONC as a
guide. \citet{reggiani2011} showed that only $\simeq 0.16$~dex of the
0.4~dex apparent age spread could be explained with observational
uncertainties of types A or B. The rest (about 80\%) we choose to
attribute to protostellar accretion histories, a case C
uncertainty. Our assumed combination of the 3 cases has only a minor
effect on the result.
As we show in Appendix \ref{appendix:simulations_simplespreads}, the
three uncertanty types produce similar biases in $\eta_{\rm meas}$,
differing only by systematic offets of a few tenths.

Lastly, in a separate set of simulations, we also include a
correction to the isochrones to remove the small mass-age correlation
introduced by using the Siess models to interpret the observed HRD (see
Appendix \ref{section:acchistory}), as this turns out to have a far larger effect on $\eta$ than even the adoption of a different family of evolutionary models (Appendix \ref{appendix:evolmodels}). When the Siess isochrones (and most
of other evolutionary models) are used to assign stellar parameters,
intermediate mass stars appear older, on average, than low mass
stars. We make the assumption that this is a systematic error and
tilt the simulated, broadened population in the HRD by the amount that
would remove any correlation between mass and age in the observed ONC
population. Equation \ref{equation:typeC} is then used to calculate a
new $\dot M_{\rm acc}$ from the resulting perturbed measured stellar parameters.

Results comparing the recovered $\eta_{\rm meas}$ versus the input $\eta_{\rm true}$ are shown in Figure \ref{fig:tautau_lumi_teff_underl}, separating the assumptions of an uncertainty in $T_{\rm eff}$ or of differential $A_V$, and including or not the correction of the isochrones for the mass-age correlation. Each simulation is performed for two input mass dependences of $\dot M_{\rm acc}$: $b=0$ (no mass dependence) and $b=2$ (large mass dependence). Results are presented for the entire mass range and also for three subsets of $M_*$. We find that $\eta_{\rm true} \simeq 3\eta_{\rm meas}$ and that when considering only the $T_{\rm eff}$ or $A_V$ uncertainties (upper panels of Fig.~\ref{fig:tautau_lumi_teff_underl}), this bias is not sensitive to $b$ and only weakly dependent on stellar mass. Thus, for the level of uncertainties we have assumed, a value of $\eta_{\rm meas}\simeq 1.4$ that is typically found in the ONC and other young Galactic clusters (see Section \ref{section:intro}) can only be reproduced if $\eta_{\rm true} \sim 4.3$, which indicates a much faster decay of $\dot M_{\rm acc}$ than commonly assumed.

It is noteworthy that the simulations with spreads in $T_{\rm eff}$ or $A_V$ {\it and} $\log L_*$ (upper panels of Figure \ref{fig:tautau_lumi_teff_underl}) yield a similar bias to those found from simulations with only a luminosity spread of the same total extent (see Figure \ref{fig:simplespread_tautau02} in Appendix~A). The reason for this is presumably that uncertainties in $T_{\rm eff}$ or $A_V$ at the level we have assumed do not lead to $\log L_*$ errors that contribute in a significant way to the observed $\log L_*$ dispersion, which is therefore dominated by the additional uncertainty in $\log L_*$ that we have inserted in the simulations.

When we correct the isochrones to eliminate a possible mass-age correlation in intermediate mass stars (see section~\ref{section:acchistory}), the bias in $\eta$ remains similar (Fig.~\ref{fig:tautau_lumi_teff_underl}, bottom panels). However, the small mass dependence seen in the upper panels is completely removed. This phenomenon is a coincidence, but simplifies any quantitative estimate of the bias in $\eta$. On the other hand the bias does now depend on $b$, getting stronger as $b$ increases.

\subsection{Changing the assumed real age spread}

We have performed similar simulations changing the assumed real age spread between $0.1 \leq \sigma \log t \leq 0.3$~dex. We have considered the same uncertainties in $T_{\rm eff}$ and $A_V$, but varied those in $\log L_*$ in order to reach the same total apparent age spread $\sigma\log t_{\rm meas}=0.4$. As before, we find the results almost indistinguishable from those derived in Appendix \ref{section:simulations-simplespread-varyingspread} when considering solely the effect of a luminosity spread; the bias in $\eta$ is very sensitive to the balance between any real age spread and the apparent age spread due to uncertainties in $\log L_*$.  Our results predict that the bias in $\eta$ varies from $\eta_{\rm true} \simeq 10\, \eta_{\rm meas}$ for $\sigma\log t=0.1$~dex, to $\eta_{\rm true} \simeq 1.7\, \eta_{\rm meas}$ for $\sigma\log t=0.3$~dex.

As anticipated, observations in a number of regions have shown systematically lower values of $\eta_{\rm meas}$ ($\lesssim1$) when the stellar samples are restricted to the solar-mass range \citep[e.g.,][]{demarchi2010,beccari2010,spezzi2012}. Our simulations accounting for all uncertainties (Figure \ref{fig:tautau_lumi_teff_underl} bottom panels) on the other hand predict a mass-independent bias in $\eta$, so that a slower decay of $\dot M_{\rm acc}$ for intermediate masses compared to low-mass stars may be genuine. However, our modelling assumes uniform uncertainties at all masses and this may not be realistic. Observations often show systematically larger apparent age spreads for solar-type stars \citep[e.g.,][]{dario2010a,dario2010b}, which could indicate that this mass range is more affected by uncertainties, thus experiencing a larger bias in $\eta$. Luminosity spreads caused by accretion history may also be more important for solar-mass objects \citep{hosokawa2011} and since this type C uncertainty produces the largest bias (see Appendix \ref{section:simulations-simplespread-ONC}), a smaller $\eta_{\rm   meas}$ might be found for intermediate-mass objects even if $\eta_{\rm true}$ is mass-independent.

\subsection{Combining uncertainties and selection effects}
\label{section:combined effects_uncertanties_seleffects}

\begin{figure*}
\includegraphics[width=1.6\columnwidth]{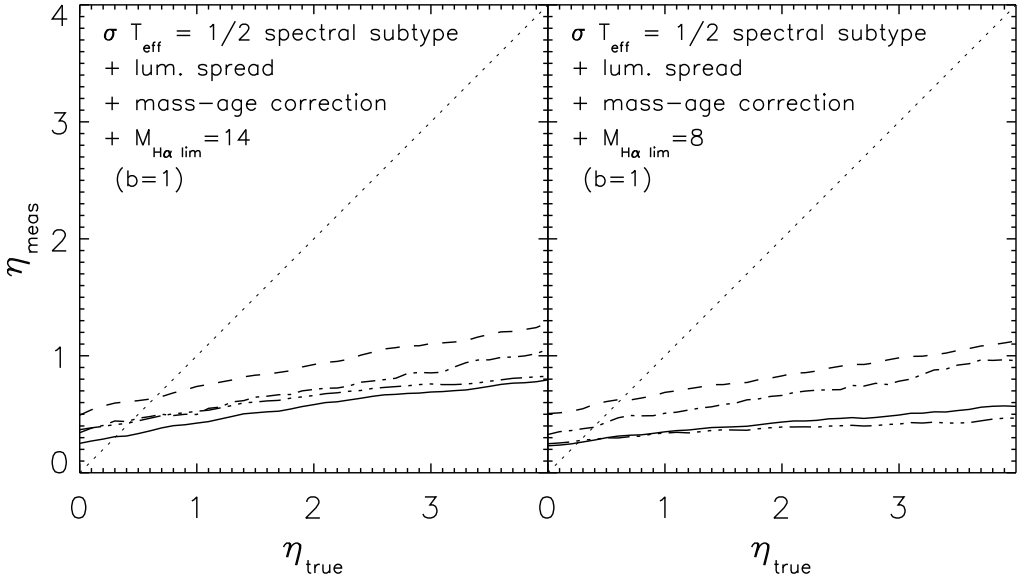}
\caption{Measured $\eta$ as a function of the real $\eta$, after accounting for both a combination of observational uncertainties (see text) and selection effects due to incompleteness, assuming two detection limits: $m_{\rm H\alpha}^{max} = 14$ (left panel) and $m_{\rm H\alpha}^{max} = 6$ (right panel). Here the assumed mass dependence is always $b=1$. Non-solid lines correspond to the result limited to 3 mass bins, and the linestyles are as in Figure \ref{fig:simplespread_tautau02}.  \label{fig:tautau_lumi_teff_underl_seleffects}}
\end{figure*}

We now add selection effects due to detection incompleteness to our simulations. A full description of the origin of the bias in $\eta$ due to this effect alone is given in Appendix \ref{section:selection_effects}. We first run a synthetic population simulation as in Section \ref{section:combined effects}, assuming an intrinsic age spread $\sigma \log t=0.2$ and the same combination of uncertainties to reach a total apparent age spread $\sigma \log t_{\rm meas}=0.4$. For simplicity, since the effects of uncertainties in $T_{\rm eff}$ and $A_V$ produced similar effects, we only include the $T_{\rm eff}$ uncertainty. The simulations include a correction to the isochrones that removes the apparent correlation between mass and age. Starting from the resulting \emph{observed} $T_{\rm eff}$ and $\log L_*$, we introduce sample incompleteness following the same method as in Appendix \ref{section:selection_effects}. This assumes that $\dot M_{\rm acc}$ is derived from H$\alpha$ photometry, which has a parameterized detection limit and photometric error distribution. Simulated sources that are a) fainter than the detection limit, or b) have H$\alpha$ excess smaller than the photometric uncertainties, are removed from the sample and excluded from the fit in the $\dot M_{\rm acc}$-$t$ plane used to derive $\eta_{\rm meas}$. Two detection limits are simulated: one relatively deep ($m_{\rm   H\alpha}^{max}=14$) and one shallow ($m_{\rm H\alpha}^{max}=8$). We only consider one value for the mass dependence of $\dot M_{\rm acc}$.  The normalization of $\dot M_{\rm acc}$ in the synthetic population modelling (see Section \ref{section:simulation_howtheywork}) leads to different fractions of excluded sources for different values of $b$ at a given detection limit. This would provide an inconsistent comparison of results for the same $\eta_{\rm true}$. Instead we consider only $b=1$, which is intermediate to the cases studied in the last section. As in Appendix \ref{section:selection_effects}, we introduce an additional dispersion of 0.65~dex in the individual values of $\dot M_{\rm acc}$, which represents star-to-star variations of accretion at a given mass and age. This value has been tailored so that, after the addition of observational uncertainties and the mass dependence of $\dot M_{\rm   acc}$, the overall dispersion from the fitted power law in the $\dot M_{\rm acc}$--$t$ plane is 0.85~dex, as found in the ONC \citep{manara2012}.

Figure \ref{fig:tautau_lumi_teff_underl_seleffects} compares $\eta_{\rm   meas}$ and $\eta_{\rm true}$ for these scenarios. By comparing these plots with those without the inclusion of selection effects (Figure \ref{fig:tautau_lumi_teff_underl}) it appears that incompleteness further displaces the $\eta_{\rm meas}$ towards unity. This is because the photometric detection threshold in the $\log\dot M_{\rm acc}$--$\log t$ plane is generally inclined with a (negative) slope $\sim1$ (see Appendix \ref{section:selection_effects}) and $\eta_{\rm meas}$ is biased towards larger values if the real $\eta_{\rm true}<1$, and towards smaller values if $\eta_{\rm true}>1$. The typical measured values $\eta_{\rm meas}\sim1.4$ (in the ONC and elsewhere) would require $\eta_{\rm true} > 4$. However the comparison between Figure \ref{fig:tautau_lumi_teff_underl_seleffects} and Figure \ref{fig:tautau_lumi_teff_underl} also shows that this latter result is still dominated by the bias caused by uncertainties in the stellar parameters alone (Figure \ref{fig:tautau_lumi_teff_underl}) and hence by the extent of the real age spread --  $\sigma\log t=0.2$ in this case. We have also checked the behavior of the results starting from a smaller or larger $\sigma \log t$. As we found in Appendix \ref{section:simulations-simplespread-varyingspread}, a smaller $\sigma \log t$ leads to larger $\eta_{\rm true}/\eta_{\rm   meas}$ and vice-versa.

\section{Discussion}
\label{section:discussion}
Our modeling has shown that uncertainties in the parameters of PMS stars and sample incompleteness strongly bias the measured correlation between accretion and age, parameterised as $\dot M_{\rm   acc}\propto t^{-\eta}$, leading to smaller values of $\eta_{\rm   meas}$ or equivalently, overestimated accretion timescales. The size of the bias in $\eta$ depends chiefly on the extent to which the apparent age spread in a region/cluster is explained by a real age spread within the region rather than various observational uncertainties and physical mechanisms that broaden the luminosity distribution in the HRD. This has two general implications:
\begin{enumerate}
\item The real decay of mass accretion is faster than that inferred from any measured correlation between $\dot M_{\rm acc}$ and isochronal ages.
\item If $\eta_{\rm true}$ is constrained, either on a theoretical
  basis, or from observational evidence that is not based on isochronal ages in a single cluster (e.g., from the measurement of the mean $\dot M_{\rm acc}$ or the fraction of accretors in clusters of different mean age), then comparison between $\eta_{\rm true}$ and $\eta_{\rm meas}$ places a constraint on the real age spread in a cluster.
\end{enumerate}

We have shown that assuming a real age spread $\sigma\log t\sim0.2$~dex combined with luminosity uncertainties, which yield a total apparent age spread of 0.4~dex \citep[e.g.,][]{jeffries2011}, leads to $\eta_{\rm   true} \gtrsim 3\, \eta_{\rm meas}$. This indicates that the decay of $\dot M_{\rm acc}$ is $\sim 3$ times faster than that inferred from a simple fit of observed $\dot M_{\rm acc}$ versus isochronal ages. Such observations generally suggest $\eta_{\rm meas} \sim1$--$1.5$ and hence $\eta_{\rm true} \sim 4$. This implies an accretion timescale ($\tau_{\rm acc}\simeq \langle t\rangle/\eta_{\rm true} \lesssim 1$~Myr), shorter than that inferred from the measured fraction of accreting sources in clusters of different age ($\tau \simeq 2.3$~Myr, \citealt{fedele2010}). This interpretation however is not without caveats: First, precise comparison of these methods is not straightforward; the fraction of accretors is defined above a given threshold (e.g., in \citealt{fedele2010} $\dot M_{\rm   acc}>10^{-11}$~M$_\odot$yr$^{-1}$), and when this fraction is small, at cluster ages of 5--10~Myr, it carries little information about the \emph{average} values of $\dot M_{\rm acc}$ in the entire population. Second, whereas $\eta$ is weakly affected by the absolute age scale (see Section \ref{section:simulations} and appendix \ref{appendix:evolmodels}) -- which is highly dependent on the evolutionary models one adopts --
$\tau_{\rm acc}$ is proportional to the assumed mean cluster age. Still, if the current age scales were systematically wrong \citep{naylor2009}, both disk and accretion lifetimes would have to be corrected in a similar way \citep{bell2013}.

%RDJ changed - check
If observational and physical uncertainties dominate the luminosity
spread observed in young clusters, such that the real (gaussian)
dispersion in $\log$ age is smaller than 0.2~dex, then or simulations
show that $\eta_{\rm true}$ would need to be even larger ($\gg4$) to
reproduce the observed $\log \dot M_{\rm acc}$ vs $\log t$
relationship. Such a rapid, almost instantaneous, decay of $\dot M_{\rm
  acc}$ would be incompatible with both observational evidence and the theory of disc evolution. If accretion ceased this rapidly within the first Myrs of evolution, clusters of older mean age ($\sim10$~Myr) should not host $any$ accreting stars, but a small fraction of accreting older stars are observed in such clusters \citep[e.g.,][]{jayawardhana2006,fedele2010}. Also, $\eta>4$ implies a decrease of the average $\dot M_{\rm acc}$ of more than 2 orders of magnitude between 1 and 3~Myr, and clusters with mean ages in this age range do not exhibit systematic differences in their average accretion rates to this extent \citep[e.g.][]{sicilia-aguilar2006}. These observations together suggest that a very fast cluster formation scenario producing a real age spread $\sigma \log t <0.2$~dex is not favoured.

On the other hand, if most of the observed spread in age were real (more than 0.2~dex dispersion in log age), the bias in $\eta$ is smaller, allowing longer accretion timescales. It is difficult to reconcile this, at least in the case of the ONC, with the lack of correlation between disc fractions and isochronal ages \citep{jeffries2011}. Thus it appears most likely that the spread in log age for the ONC should not be far from 0.2~dex to satisfy these constraints. Assuming a mean age for the ONC of 2.5~Myr and a lognormal age distribution, this implies that 95\% of stars were born between 1.3 and 4~Myr ago. This is of order 5-10 crossing times in the region \citep[e.g.,][]{hillenbrand1998}, indicative of a relatively slow star formation process. Considering therefore $\sigma\log t=0.2$, and the typical values of $\eta_{\rm meas}$, our simulations suggest $\eta_{\rm true}\gtrsim4$.

The fast decay of $\dot M_{\rm acc}$ implied by our results is not
predicted by models that assume viscous disk evolution, \citep[$\eta
  \sim 1.5$,][]{hartmann1998}. This indicates that other factors such
as photoevaporation and high energy photons from the central star
\citep{pascucci2009,pascucci2012,owen2013} or planet formation may
either accelerate the clearing of material in the inner disk during the
first few Myrs of PMS evolution, or inhibit its accretion onto the
stellar surface. The measured fraction of sources showing warm dust
emission excess \citep[e.g.,][]{haisch2001,hernandez2008} in different
young clusters may point to longer timescales for inner disk
dissipation than the accretion timescale our finding suggest. There is
also observational evidence for fast inside-out dispersal
\citep{koepferl2012}, which supports the notion that the accretion
timescales is shorter than the disk dispersal timescale.

\subsection{Possible factors that mitigate the result}
Our analysis shows that the apparent flattening in $\eta$ is driven mainly by the fraction of the observed age spread that can be attributed to luminosity dispersion caused by either observational or physical uncertainties. The simplest solution to mitigate the requirement for a very fast decay of $\dot M_{\rm acc}$ is to allow this fraction to be small and hence a large real age spread.  However, there are other possibilities. We have made the general assumption that $\dot M_{\rm acc}$ depends solely on $t$ and $M_*$, with only an additional possible scatter (e.g., Section \ref{section:combined   effects_uncertanties_seleffects}). In fact this is not the case: $\dot M_{\rm acc}$ is also affected by other parameters, such as the temperature, mass and structure of the disc.

Let us consider a change in stellar radius produced by protostellar
history; our type C uncertainties. We assume that the temperature of
the inner disk is dominated by irradiation from the central object,
which is appropriate for low (present-day) $\dot M_{\rm acc}$. If a
star shrinks rapidly due to an intense previous accretion phase, the
disk temperature $T_D$ decreases $\propto R^{\frac{3}{4}}$. According
to models of viscous disk evolution \citep[e.g.,][]{hartmann1998}, this
affects the disk viscosity, and $\dot M_{\rm acc}\propto T_D \propto
R^{\frac{3}{4}}$. Since (at constant $T_{\rm eff}$) the isochronal age
$t\propto L^{-\frac{3}{2}}\propto R^{-3}$, we have that $\dot M_{\rm
  acc} \propto t^{-\frac{1}{4}}$, where the age $t$ here is not the
real age, but that inferred from the HRD. Thus if a star appears more
compact so that its age is overestimated by 1 dex, its $\dot M_{\rm
  acc}$ is expected to be lower by $\sim 0.25$~dex. However, this is a
relatively small effect compared to the major biases we have found in
this paper. In particular it seems incapable of mitigating the
conclusion that age spreads $\sigma \log t <0.2$ are unlikely.
 For a coeval population where the apparent age spread is entirely due
 to the protostellar accretion history $\eta_{\rm meas}$ would be
 $-0.5\lesssim\eta_{\rm meas} \lesssim0.1$ (see Appendix \ref{section:simulations-simplespread-coeval} for type C), and hence much smaller than observed in young clusters.

Another possibility is that stars which have undergone vigorous protostellar accretion have current disk properties that inhibit further accretion. For example, their discs could have a lower inner temperature, be less massive, be more stable, or be truncated at a larger inner radius. In the scenario of \citet{baraffe2012}, PMS stars may accrete their mass through short-lived intense accretion bursts, reaching up to $10^{-4}$~M$_{\odot}$\,yr$^{-1}$.  During the burst phase, the inner disk heats up to very high temperatures ($10^5$~K, \citealt{zhu2009}), driving intense accretion through MRI. If most of the stellar mass is built up by this process in a relatively short time ($<1$~Myr), it is possible that at later stages $\dot M_{\rm acc}$ remains on average lower than a star with a more steady and quiescent accretion history. Thus, a fraction of ``bursting'' stars appear underluminous -- hence, older -- and with lower $\dot M_{\rm acc}$, mimicking an apparent decay of accretion even though they are coeval with the rest of the population.
This speculative scenario may allow young populations to be more coeval
than inferred from the HRD and for their accretion timescales to be longer. However it is completely unclear to what extent present-day and historical accretion rates are anti-correlated (or correlated). It would take a large systematic difference in the present day $\dot M_{\rm acc}$, of 2 or more orders of magnitude between stars with different protostellar accretion histories to compensate for our finding that $\eta\gtrsim4$ if $\sigma \log t \simeq  0.2$~dex in Orion.

Last, as we anticipated in Section \ref{section:the_problem_mdot}, the standard methods to derive $M_{\rm acc}$ from $L_{\rm acc}$, i.e., Equation \ref{equation:mdot1}, are a simplification of a more complex process. In particular the assumption $R_{in}\sim5 R_*$ is not realistic, as the inner disk radius is close to the co-rotation radius. If rotation rates are independent of isochronal ages, $R_{in}/R_*$ increases with isochronal age as the star shrinks. In addition, in the ONC \citet{littlefair2011} measured systematically slower rotation rates for older stars, so that their $R_{in}$ would be even larger. For the same measured $L_{\rm acc}$, a larger $R_{in}$ implies a smaller $\dot M_{\rm acc}$ than that derived from Equation \ref{equation:mdot1}. Thus, correcting for this effect would result in a larger $\eta_{\rm meas}$ even {\em before} correcting for the biases treated in this paper, making the main conclusions of this work stronger. Also, as we discussed in Section \ref{section:the_problem} and \ref{section:simulations}, the interplay between stellar properties and evolution, accretion process, and disk properties may involve more complicated correlations which could further modify the actual bias in $\eta$; however, we find it unlikely that additional unknown effects could forcibly invalidate our main conclusions.

\subsection{General validity}

This work is based on simulations that aim to represent observational
results in the ONC. As we discussed in Section~3.1, the spread of
luminosities at a given $T_{\rm eff}$ in the ONC is quite typical of
other star forming regions.

Our simulations adopt a particular family of evolutionary models, the \citet{siess2000} isochrones. \citet{manara2012}  showed that assuming 3 different families of evolutionary models to assign stellar parameters and $\dot M_{\rm acc}$--$t$ values, $\eta$ and $b$ remain nearly the same even though the average age of the cluster is model dependent by up to a factor 2 \citep{hillenbrand2008,reggiani2011}. In Appendix \ref{appendix:evolmodels} we further demonstrate that also the bias in $\eta$, as the ratio $\eta_{\rm true}/\eta_{\rm meas}$ is very weakly affected by the choice of evolutionary models, as it turns out to be nearly identical the simulations adopt \citet{dantona1998} isochrones rather than Siess models, even though the two grids differ by more than a factor 2 in their absolute age scale. Thus our particular choice of evolutionary models is largely irrelevant to our results.

Our analysis is based on a single IMF \citep{kroupa2001}. Although
large deviations from standard IMFs are generally not observed for
young clusters, at least in the stellar mass range \citep{bastian2010},
in principle a different IMF might affect our results. However, our
simulations show, that the bias in $\eta$ is very similar even when analysis
is restricted to specific mass ranges. Thus reasonable IMF variations would not affect our results.

There are no reasons to suppose that the accretion timescales in the ONC are unrepresentative of typical PMS evolution:  $\eta_{\rm meas}$ in the ONC is comparable with values measured with similar techniques (stellar parameters derived with spectroscopy, $\dot M_{\rm acc}$ from measuring flux excesses and Equation \ref{equation:mdot1}) in other regions \citep{hartmann1998,muzerolle2003,muzerolle2005,calvet2004,mohanty2005,natta2006,garcialopez2006,gatti2008,biazzo2012,fang2009,fang2013}. Since these analyses share similar methods, and given that most of the uncertainties affecting young stars are either intrinsic to the nature of these objects, or due to the simplifying assumptions used to derive $\dot M_{\rm acc}$ from observed quantities, we see no reasons to assume that uncertainties in the ONC have a peculiar amplitude.

%RDJ changed - check
There remains the possibility that something in the ONC environment
could affect the evolution of accretion rates or the extent of any real
age spread. O-stars in the central part of the ONC are observed to be
externally photoevaporating disks, yet the fraction of members with
disks and detectable accretion ($\sim 2/3$, \citealt{jeffries2011}) is
similar to lower mass clusters at the same age
\citep{hernandez2008}. Dynamical interactions in the densest parts of
the ONC may have slightly affected the disk radii distribution
\citep{dejuanovelar2012}, but this also has little impact on the inner disks. As for the real age spread, it is generally assumed to depend on the cloud density \citep{krumholz2007,parmentier2013}, through the dynamical collapse time. The ONC can be considered an ``average density'' star forming regions, between compact starburst clusters and sparse young populations.

Our analysis has made the assumptions that $\eta$ is obtained from the measured $\dot M_{\rm acc}$ and isochronal $t$ in one cluster; in some cases the stellar samples considered may include multiple separate stellar populations. Under these circumstances, we distinguish two cases. If the mean ages of the populations are similar, differing by say less than their age spreads, then they can be treated as a single cluster and the bias in $\eta$ will be similar. If there is a larger difference in age, the sample could be considered as a single population with a larger real age spread (and possibly also a larger overall apparent age spread than that measured in individual clusters). The quantitative effect on $\eta_{\rm meas}$ will depend on the exact values of these age spreads; as the ratio between the overall real age spread (from multiple populations) and the total apparent observed age spread increases, the bias in $\eta$ is smaller, and vice-versa.

\section{Summary}
\label{section:summary}

%RDJ changed - check
In this work we have investigated how uncertainties in estimated stellar
parameters (mass, radius, age) and observational selection effects bias estimates of the temporal decay of
mass accretion rates, parameterised as $\dot M_{\rm acc}\propto
t^{-\eta}$. These uncertainties may be due to observational limitations
(e.g. photometric errors, differential extinction, accretion veiling
etc.) or to physical mechanisms such as variations in their
protostellar accretion histories \citep{baraffe2012}.  The overall
extent of this broadening is not well known, but many studies have
suggested that these uncertainties are not negligible. As a consequence
the bias in the measured value of $\eta$ originates from a combination
of attenuation bias in the linear regression of $\log \dot M_{\rm
  acc}=-\eta \log t\, +$~const -- since  $\log t$ has large relative
errors -- and correlated uncertainties between the estimated stellar parameters and $\dot M_{\rm acc}$.

Our analysis assumes $\dot M_{\rm acc}$ depends only on age and
mass. We have performed Monte Carlo simulations that introduce
uncertainties into the Hertzsprung-Russell diagram and observational
selection effects and realistic combinations of these.
In all cases the recovered $\eta_{\rm meas}$ inferred from the
observational data is smaller than $\eta_{\rm true}$. If multiple types
of uncertainties are simultaneously added, the relative contribution of
each is largely irrelevant; the result is dictated by the overall
extent of uncertainties, which in turn depend on the {\it apparent} age spread determined from placing objects in the HRD and the fraction of this that is due to a {\it real} spread in age.

Assuming a typical apparent age spread observed in young clusters
of $\sigma \log t=0.4$~dex but that only 0.2~dex of this is attributable
to a real age spread
(for example in the Orion Nebula Cluster,
\citealt[e.g.,][]{jeffries2011}), then  the $\eta_{\rm meas}\sim$1--1.5
that is typically measured in the ONC and other young Galactic
clusters, can only be reproduced if $\eta_{\rm true}\gtrsim 4$.
If the real age spread were smaller ($\sigma \log t<
0.2$~dex), as advocated by fast star formation scenarios, the bias
becomes extreme, implying an almost instantaneous decay of $\dot M_{\rm
  acc}$, which is incompatible with the detection of accretors in
young clusters over a wide range of mean ages (1-10~Myr).
We conclude that clusters like the ONC cannot be coeval and must have
an age spread $\sigma \log t \geq 0.2$~dex. Conversely,
assuming the minimum levels of observational uncertainty, the
corresponding real age spread be must be $\sigma \log t \leq
0.3$~dex, and means $\eta_{\rm true}$ would still be at least twice $\eta_{\rm meas}$.
The present uncertainties in the absolute age scale of PMS stars and
young clusters do not allow us to obtain an accretion decay timescale;
however the bias in $\eta$ is only weakly affected by the assumed age
scale.

Observational selection effects, leading to incomplete samples as a
function of mass or age introduce an additional bias to
$\eta_{\rm meas}$. Depending on the detection limit and intrinsic
scatter in the accretion rates, $\eta_{\rm meas}$ is always biased
towards $\sim$1. Thus, a combination of selection effects together with
uncertainties in the stellar parameters will worsen the apparent
flattening of $\eta$ when $\eta_{\rm true}>1$, strengthening our
conclusion that the decay of $\dot M_{\rm acc}$ is much faster than
the observed relationship between $\log \dot M_{\rm acc}$ and $\log t$.

This result, and the suggestion that the real age spread in a young
cluster cannot be very small, could be mitigated if there were a strong
anti-correlation between present-day $\dot M_{\rm acc}$ and any
previous protostellar accretion that gave the star a smaller radius and
the {\it appearance} of being older. Conversely, correcting the
simplistic assumption that the inner disk radius is proportional to the
stellar radius, by allowing for the influence of stellar rotation on
disk truncation, could exacerbate the bias. Additional scenarios
originating from a more complex interplay between stellar, disk and
accretion properties as well as observational biases in their
estimation may add further bias in $\eta$ or partially mitigate our
results. However, the magnitude of the bias in $\eta$ we have found
is so large that unknown additional effects would have
to introduce very strong correlations to alter our main conclusions.

Lastly, our analysis cannot directly explain the observation that the
measured $\dot M_{\rm acc}$--$t$ relation can appear flatter for
solar-mass PMS stars than those of lower mass
\citep{manara2012,demarchi2011b,demarchi2012}. One possibility is that
any radius and hence apparent age perturbation due to varying
protostellar accretion histories, plays a more important role, and
introduces a greater bias in the measured $\eta$, in intermediate mass stars, as suggested by \citet{hosokawa2011}.

\noindent{\em Acknowledgments}
\newline
\noindent We thank the anonymous referee for helpful comments that improved the manuscript, and Guido De Marchi for helpful discussions on the interpretation of the results. 

%%%%%%%%%%%%%%%%%%%%%%%%%%%%%%%%%%%%%%%%%%%%%%%%%%%%%%%%%%%%%%
%%%%%%%%%%%%%%%%%%%%%%%%%%%%%%%%%%%%%%%%%%%%%%%%%%%%%%%%%%%%%%
%%%%%%%%%%%%%%%%%%%%%%%%%%%%%%%%%%%%%%%%%%%%%%%%%%%%%%%%%%%%%%
%%%%%%%%%%%%%%%%%%%%%%%%%%%%%%%%%%%%%%%%%%%%%%%%%%%%%%%%%%%%%%

%%%%%%%%%%%%%%%%%%%%%%%%%%%%%%%%%%%%%%%%%%%%%%%%%%%%%%%%%%%%%%
%%%%%%%%%%%%%%%%%%%%%%%%%%%%%%%%%%%%%%%%%%%%%%%%%%%%%%%%%%%%%%
%%%%%%%%%%%%%%%%%%%%%%%%%%%%%%%%%%%%%%%%%%%%%%%%%%%%%%%%%%%%%%
%%%%%%%%%%%%%%%%%%%%%%%%%%%%%%%%%%%%%%%%%%%%%%%%%%%%%%%%%%%%%%

\appendix
\section{$\dot M_{\rm acc}$ vs $t$ biases from individual uncertainties in the HRD}
\label{appendix:simulations_simplespreads}

This appendix describes in the detail the simulations of synthetic populations to analyse the bias in the $\dot M_{\rm acc}$ temporal decay exponent $\eta$ due to uncertainties in the stellar parameters. We separately treat the effect of luminosity spread, $T_{\rm eff}$ and $A_V$ errors, and systematic mass-age correlations due to inaccurate stellar evolutionary models.

\subsection{Apparent luminosity spread}
\label{section:simulations-simplespread}
\subsubsection{An average age spread}
\label{section:simulations-simplespread-ONC}

\begin{figure*}
\includegraphics[height=2\columnwidth,angle=90]{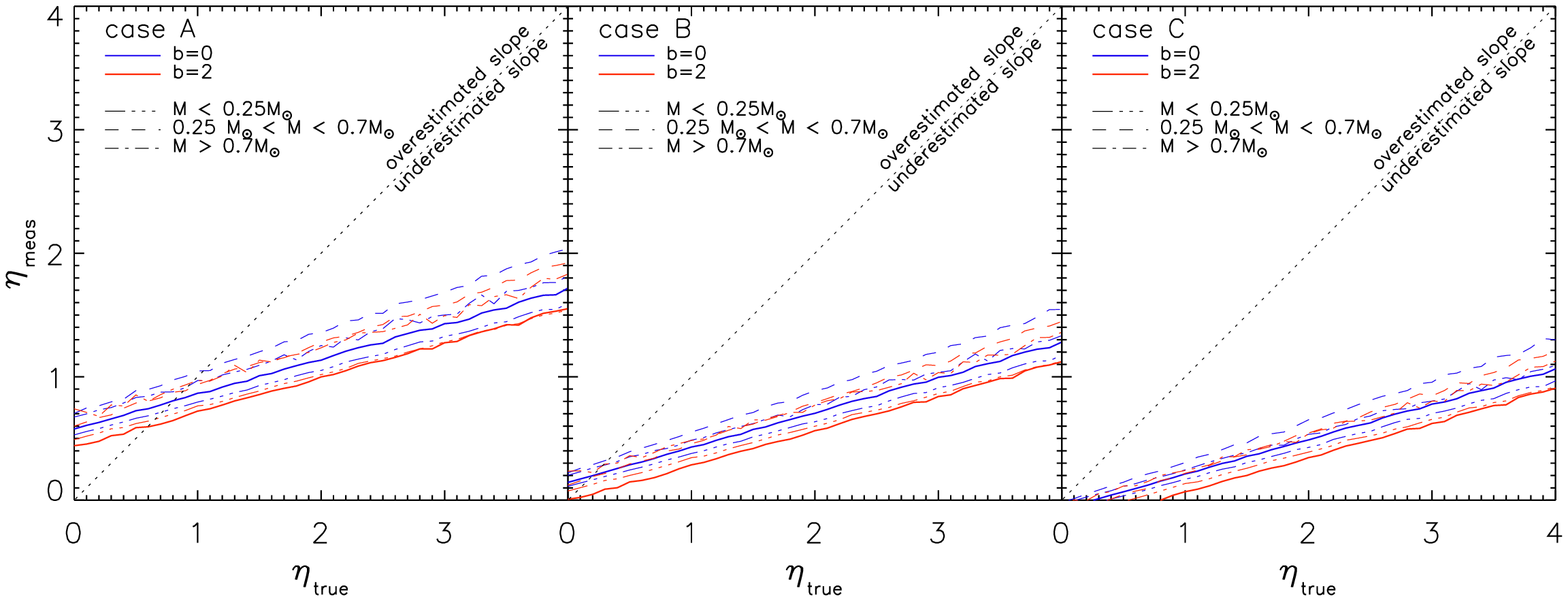}
\caption{Measured value of $\eta$ as a function of the real one, after
  adding a luminosity spread with $\sigma \log L= 0.2$ to a simulated
  2.5~Myr population with true age spread $\sigma \log t_{\rm true} =0.2$~dex. The 3 panels correspond to the limiting cases as described in the text. The red and blue lines are for two input mass dependences of $\dot M_{\rm acc}$ ($b=0$ and $2$). Results limiting the fit to specific mass ranges are shown as reported in the legend. \label{fig:simplespread_tautau02}}
\end{figure*}
We study the bias in $\eta$ introduced by an apparent luminosity spread.
We start with a simulated population with an intrinsic age spread
$\sigma\log t_{\rm true}=0.2$. This corresponds to a lognormal age distribution
peaked at 2.5~Myr and having 95\% of the stars between 1 and 6~Myr
old. This is the upper limit on the age spread in the ONC deduced by
\citet{jeffries2011} from the absence of correlations between disk
fraction and apparent isochronal age. We then apply a luminosity spread
$\sigma_{\rm errors} \log L_*=0.23$ to the simulated population; this
value is chosen to produce an overall apparent age spread $\sigma\log
t_{\rm meas}=0.4$ as observed in the ONC (e.g., Da Rio et al. 2010a).

We iterate our simulations for different assumed $\eta_{\rm true}$ and
$b$; the results for biases of types A, B and C (see Section \ref{section:simulation_howtheywork} are shown in Figure \ref{fig:simplespread_tautau02} in terms of $\eta_{\rm meas}$  versus $\eta_{\rm true}$.
This shows that the assumed luminosity spread leads to significant underestimation
of $\eta_{\rm true}$ in most cases. The value $\eta_{\rm meas}=1.4$
derived in the ONC \citep{manara2012} is recovered if $\eta_{\rm true}
\simeq $3 for uncertainties of type A, and up to 4 or more for types B
and C. Thus, assuming that only a spread in luminosities affects the
observed HRD, then $\dot M_{\rm acc}$ decreases by 3-4 orders of
magnitude between 0.5~Myr and 5~Myr.

Figure \ref{fig:simplespread_tautau02} also shows that this result is
only weakly dependent on the mass dependence of $\dot M_{\rm acc}$, as
$\eta_{\rm meas}$ changes by $\leq 0.2$ between $b=0$ to $b=2$. We find
that $\eta_{\rm meas}$ is slightly smaller for very low mass stars than for higher stellar masses. The best fit for the entire population (solid lines) generally follows the results for the lowest masses, as these represent the dominant population because of the IMF.

If we assume $\eta_{\rm true}\sim0$, which indicates a constant $\dot
M_{\rm}$ over age, the luminosity spread leads to $\eta_{\rm meas}
\neq0$; it is positive (a decay of $\dot M_{\rm}$), $\sim0.5$ for
biases of type A and $\sim 0.1$ for type B, but negative (an apparent
increase of $\dot M_{\rm acc}$ with age) for type C. Although assuming
$\eta_{\rm true}=0$ is unphysical, provided that the population is not
coeval and since $\dot M_{\rm acc}$ must decrease (in a population) over time, this offset reveals the effect of the correlation of uncertainties we discussed in Section \ref{section:intro}.

\subsubsection{Varying the age spread}
\label{section:simulations-simplespread-varyingspread}

\begin{figure*}
\includegraphics[height=2\columnwidth,angle=90]{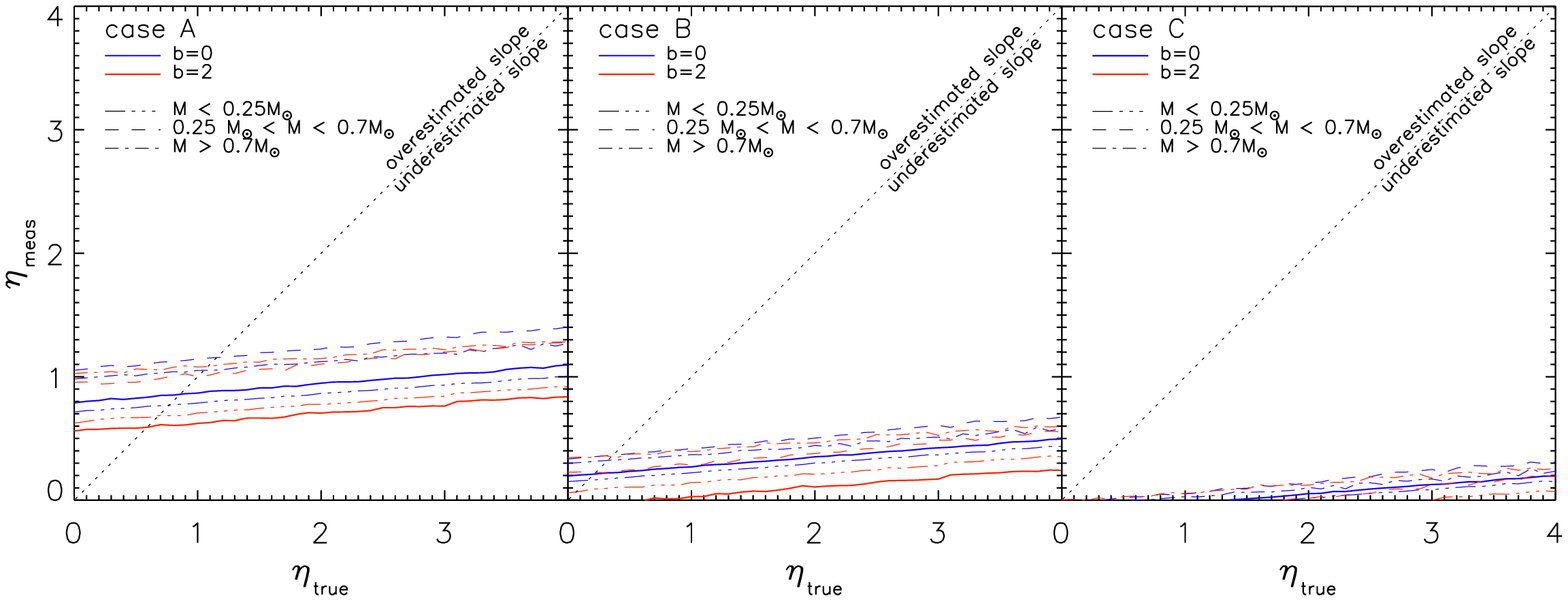}
\includegraphics[height=2\columnwidth,angle=90]{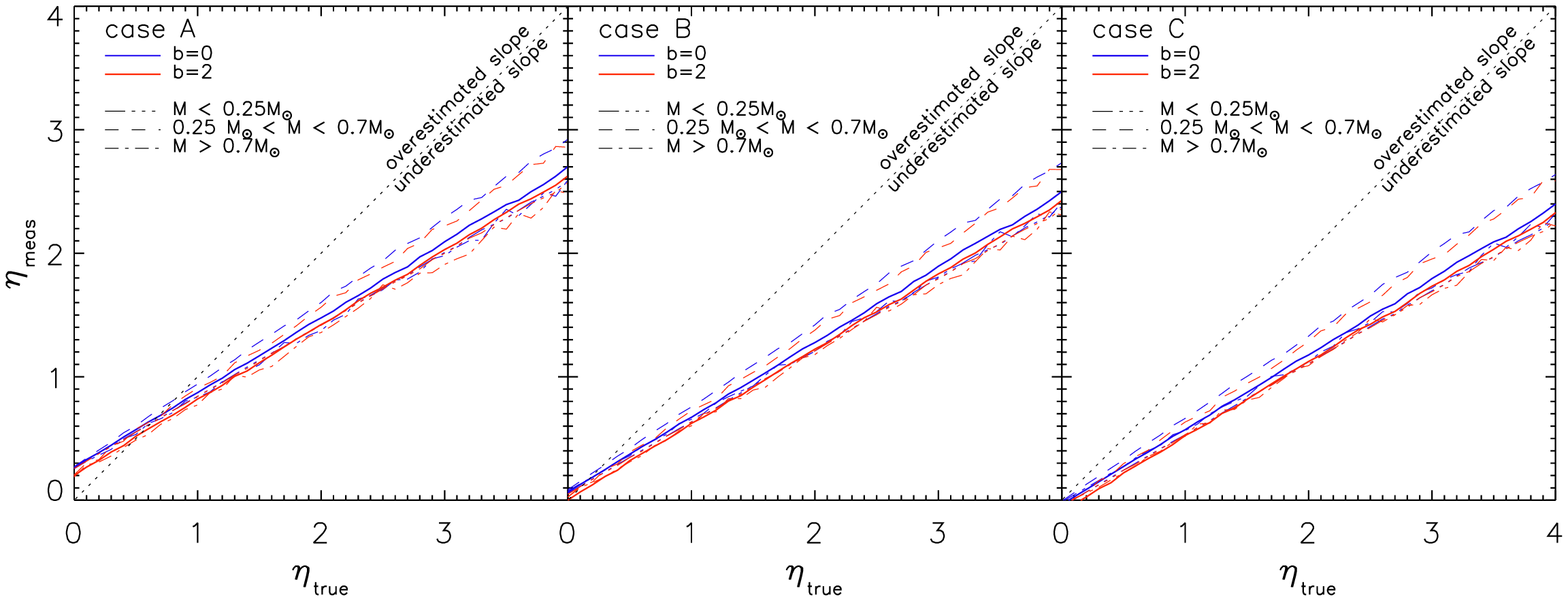}
\caption{Same as Figure \ref{fig:simplespread_tautau02}, i.e. for an
  overall apparent age spread of 0.4~dex induced by luminosity
  uncertainties, but assuming that the real intrinsic age spread is
  smaller( $\sigma \log t_{\rm true}=0.1$ upper panels) or larger ($\sigma \log
  t_{\rm true}=0.3$ bottom panels). \label{fig:simplespread_tautau013}}
\end{figure*}

We now test how the bias in $\eta_{\rm meas}$ is influenced by the
assumed real age spread $\sigma\log t$. Specifically, we test
smaller and larger values than previously assumed, namely, $\sigma
\log t_{\rm true}=0.1$~dex and $0.3$~dex. Once again, the age distribution peaks at
2.5~Myr. The simulations are as in Section
\ref{section:simulations-simplespread-ONC}, except that, in order to
reach the same apparent observed $\sigma\log t_{\rm meas}=0.4$, we add
luminosity spreads $\sigma_{\rm errors} \log L_*$ =0.27~dex and 0.15~dex for the
cases of $\sigma\log t_{\rm true}=0.1$~dex and 0.3~dex respectively.

The results are shown in Figure \ref{fig:simplespread_tautau013}. If the assumed real age spread is smaller, the bias in $\eta$ increases, producing both a more prominent flattening of the $\eta_{\rm meas}$ versus $\eta_{\rm true}$ correlation, and a larger offset from correlated uncertainties (the intercept $\eta=0$).
Such increased bias is due to the larger amount of luminosity scatter needed to obtain the same observed $\sigma \log t_{\rm meas}$ starting from a smaller $\sigma \log t_{\rm true}$. The opposite occurs when the real age spread increases and the relative contribution from apparent luminosity spreads is smaller.

In the presence of only luminosity spreads, the assumption of
$\sigma\log t_{\rm true}=0.1$ leads to a very severe bias: considering the full
mass range (solid lines), $\eta_{\rm meas}$ is always $<1.1$, even for
$\eta_{\rm true} =4$, $b=2$ and bias type A. Since in reality a combination of
cases A, B, and C is expected, a value $\eta_{\rm true} \sim 10$ or more would be needed to reproduce the slope observed in the ONC and other young regions \citep[][, $\eta\sim1.4$]{manara2012}. With such an instantaneous decay of $\dot M_{\rm acc}$, no accreting sources would be detected in clusters older than a few Myr.

\subsubsection{A coeval population}
\label{section:simulations-simplespread-coeval}

\begin{figure*}
\includegraphics[height=2\columnwidth,angle=90]{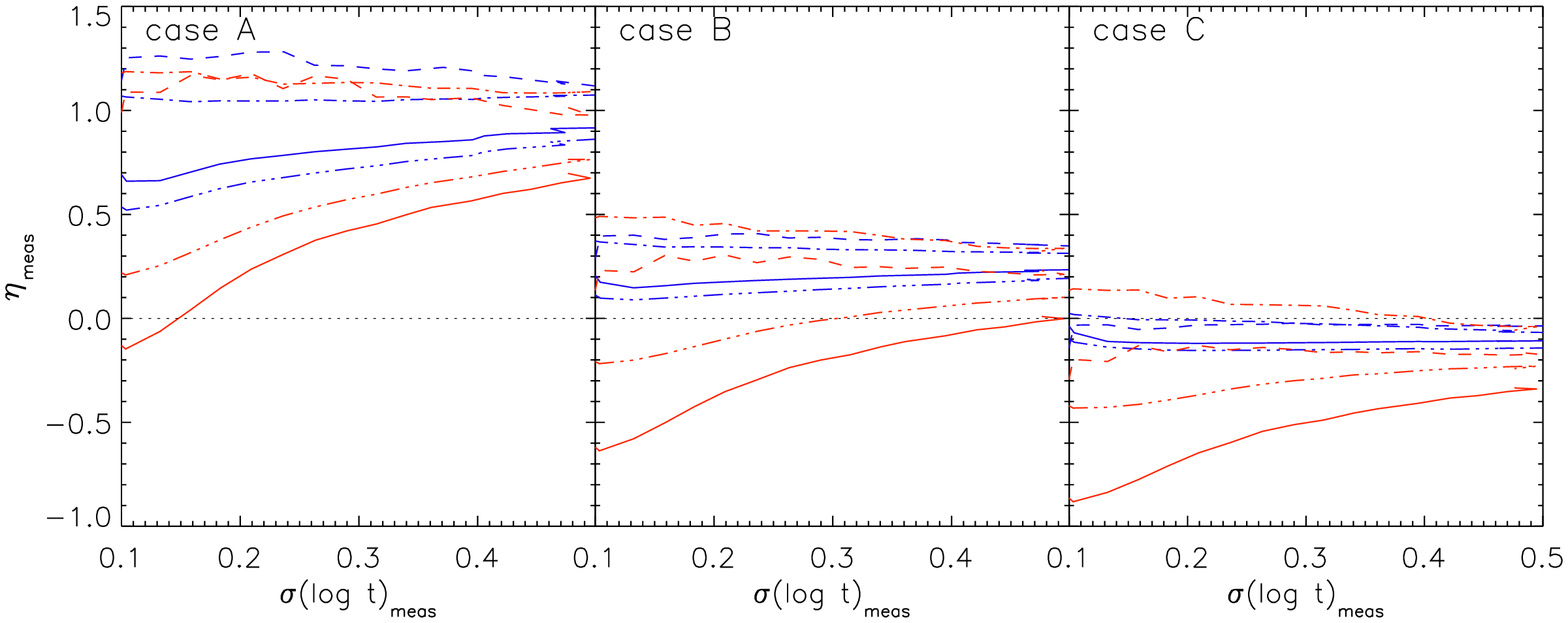}
\caption{Measured $\eta$ as a function of the observed isochronal age spread, starting from a coeval population to which a luminosity spread is added. Colors, and line styles are as in Figure \ref{fig:simplespread_tautau02}. \label{fig:simplespread_coeval}}
\end{figure*}

We also explore the limiting case of a fully coeval young stellar population, whose apparent age spread from the measured HRD is solely due to uncertainties in $\log L_*$ (bias of types A and B), or a luminosity spread induced by differences in the protostellar accretion history (bias of type C).
We start with a $2.5$~Myr population and add a luminosity spread as in the previous subsections.
The input value $\eta_{\rm true}$ is now meaningless, as all simulated
stars share the same age; instead, we vary the amount of luminosity
spread and each time we fit the measured $\log \dot M_{\rm acc}$ -- $t$
relation determining $\eta_{\rm meas}$. Again, we simulate $b=0$ or
$b=2$, and study separately the 3 types of bias A, B and C. The results
are shown in Figure \ref{fig:simplespread_coeval}, in terms of
$\eta_{\rm meas}$ as a function of the apparent isochronal age
spread. The general trend is qualitatively close to that obtained for a
small intrinsic age spread (e.g., $\sigma \log t_{\rm true}=0.1$, see
Figure \ref{fig:simplespread_tautau013} upper panel) and $\eta_{\rm
  true}=0$. For bias of type A, correlated uncertainties generate a
positive value of $\eta_{\rm meas}$, i.e., an apparent temporal
decrease of $\dot M_{\rm acc}$, with $\eta_{\rm meas}=0.5$--$0.8$ for
0.4~dex of apparent age spread and depending on the assumed $b$. On the
contrary, bias of type C produces a negative $\eta_{\rm meas}$, down to
$\sim-0.5$, and type B lies in between. As the luminosity spread will
likely be some combination of the 3 cases, then accretion rates in a
coeval population should show an apparent correlation with age with
$0\simeq\eta_{\rm meas}\lesssim0.5$.
This result, leaving aside the additional caveats reported in Section
\ref{section:discussion}, provides a test for the coevality of any
young population: it is difficult to advocate coevality for a cluster
in which $\eta_{\rm meas}\gtrsim0.5$.

\subsection{$T_{\rm eff}$ uncertainty}
\label{section:simulations-teff}
\begin{figure*}
\includegraphics[width=1.5\columnwidth]{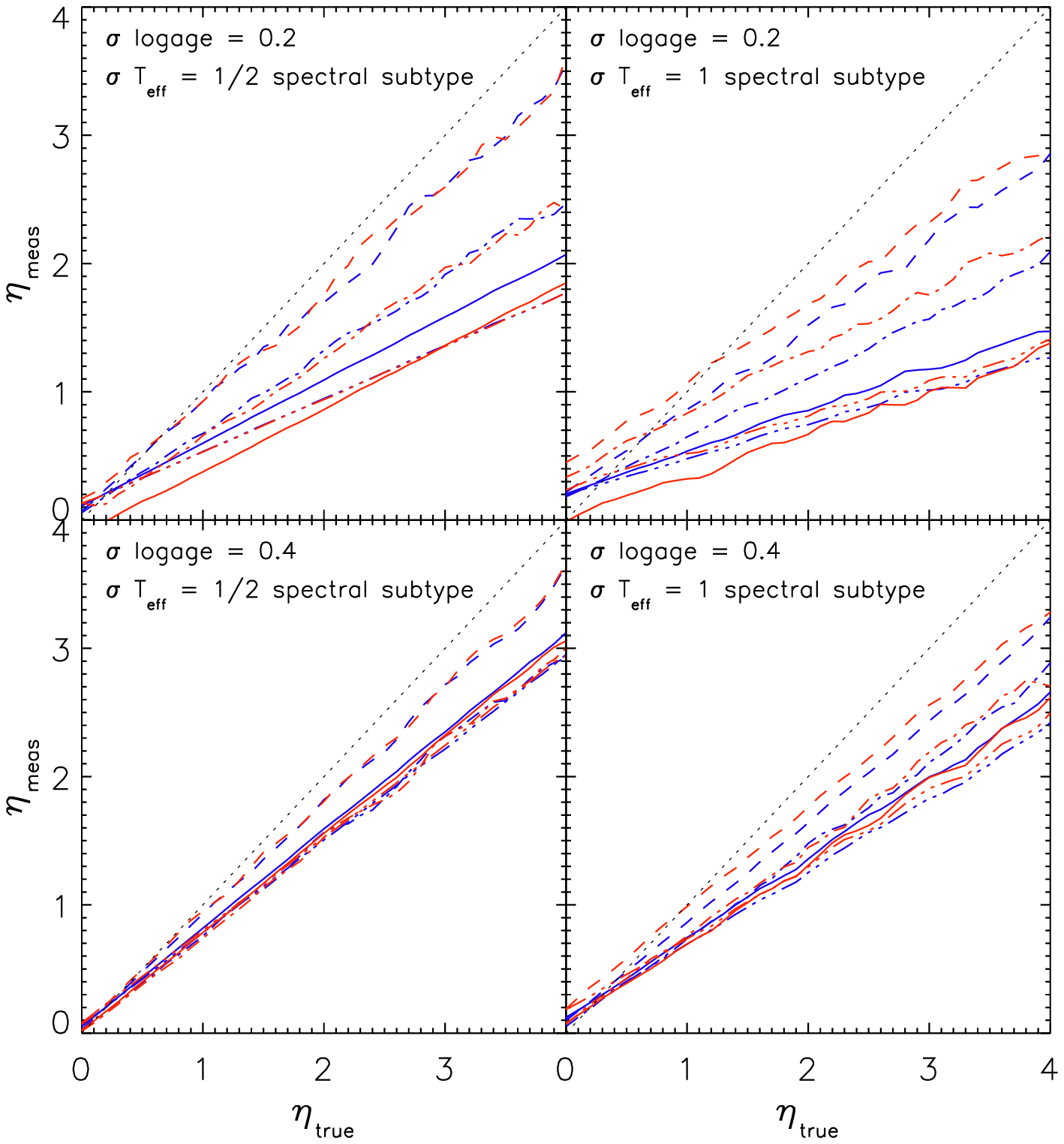}
\caption{Measured versus true value of $\eta$, after introducing an
  uncertainty in $T_{\rm eff}$ of half or 1 spectral subtype, and
  assuming an intrinsic age spread $\sigma\log t_{\rm true}=0.2$ or 0.4 dex. Colors and line styles are as in Figure \ref{fig:simplespread_tautau02}. \label{fig:teff_tautau}}
\end{figure*}

We analyze now the bias in $\eta$ from uncertainties in $T_{\rm eff}$ alone, using the same technique we adopted for spreads in luminosity (Section \ref{section:simulations-simplespread}).
We start with two values for the real age spread, $\sigma\log t_{\rm
  true}=0.2$ and $\sigma\log t_{\rm true}=0.4$.
The $T_{\rm eff}$ of the simulated stars are displaced with gaussian
shifts with standard deviation of either half or one spectral subtypes;
we assume the temperature scale of \citet{luhman1999} for the
conversion between spectral types and $T_{\rm eff}$. In principle an
error in an assigned spectral types may propagate to the estimated
$\log L_*$ as well, in two cases: a) when it leads to an error in the
estimated $A_V$, this case will be explored in Section
\ref{section:simulations-teffav}; b) when the bolometric corrections
used to obtain $L_*$ are quite sensitive to $T_{\rm eff}$. The latter
is generally not a significant effect, when $L_*$ is derived from
apparent magnitudes close to the peak of the stellar SEDs, e.g., photometry in the red optical range for low-mass stars.

We consider that the bias in $\dot M_{\rm}$ from $T_{\rm eff}$ errors always follows type A (see Section \ref{section:simulation_howtheywork}). This is a fair approximation, since, generally, errors in $T_{\rm eff}$ do not significantly affect the accretion luminosity contrast $L_{\rm acc}/L_*$. Strictly speaking, this holds if the bolometric correction of the photospheric flux, at the wavelength of the indicator used to estimate $L_{\rm acc}$ (e.g., H$\alpha$), is weakly dependent on $T_{\rm eff}$.

Figure \ref{fig:teff_tautau} shows the results: as for the luminosity spreads, $\eta_{\rm meas}< \eta_{\rm true}$. The bias is significant even for the modest error of 1/2 a spectral subtype, which is practically the smallest uncertainty achievable with spectroscopy: the flattening is at least $\sim25\%$ when the entire mass range is considered (solid lines) and starting from a large initial age spread. For larger age spreads the bias is smaller, because the stellar population is already well spread in the HRD, and thus an additional uncertainty in $T_{\rm eff}$ introduces a smaller relative broadening of the width of the age distribution.

\subsection{Differential $A_V$}
\label{section:simulations-teffav}
When the parameters of PMS stars are derived from optical data in the presence of differential $A_V$, an error in $T_{\rm eff}$ also to leads to a correlated error in $A_V$, and thus in $L_*$. This is because, given an observed color for a star, if $T_{\rm eff}$ is overestimated (e.g., from inaccurate spectral type assignment), the intrinsic photospheric color will be underestimated, the reddening overestimated, and the intrinsic luminosity also overestimated.
Thus, $T_{\rm eff}$ errors tend to produce diagonal shifts in the HRD, whose inclination depends on the photometric bands used to derive stellar parameters, and typically will be steeper for shorter wavelengths.
%RDJ changed - check
This type of error will also apply to any analyses where a uniform
reddening is assumed for a population where some level of differential
extinction exists.

\begin{figure*}
\includegraphics[width=1.5\columnwidth]{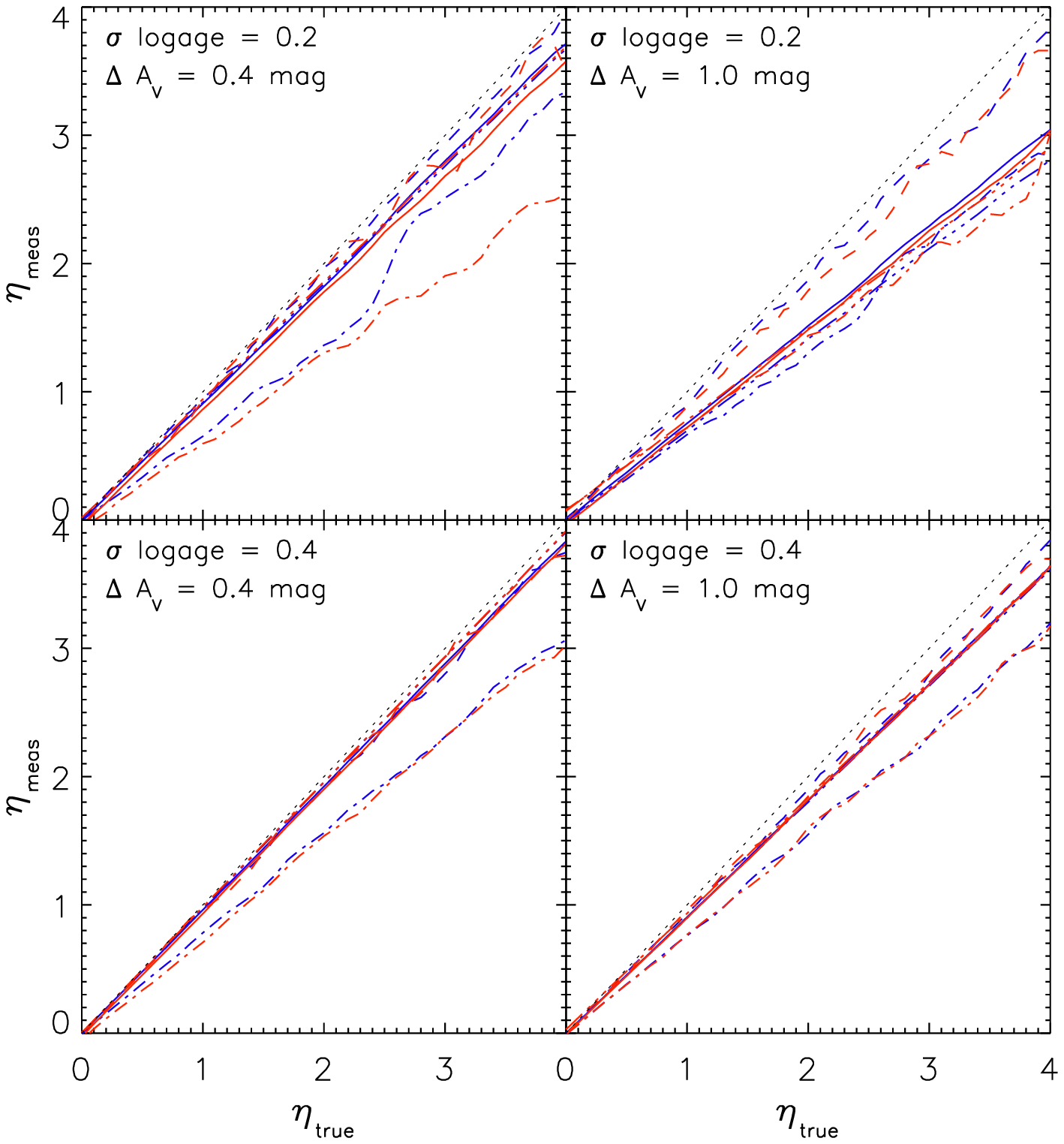}
\caption{Measured $\eta$ as a function of the real one, obtained with spreading a 2.5~Myr population in an HRD with differential reddening. Colors and line styles are like in Figure \ref{fig:simplespread_tautau02}. The 4 panels correspond to our 2 assumptions on $\Delta A_V$ and 2 initial age spreads.  \label{fig:teffav_tautau}}
\end{figure*}

We run another set of simulations as before, and apply a spread in the HRD due to extinction uncertainties. To define the correlation between $\log T_{\rm eff}$ and $\log L_*$, we assume that the HRD is derived from an optical $I$ versus $(V-I)$ CMD. First, considering the Galactic reddening law of \citet{cardelli1989}, we derive the $(V-I)$ and $I$ shifts due to $A_V=1$. These are then converted into $T_{\rm eff}$ and $\log L_*$ shifts, as a function of $T_{\rm eff}$, based on synthetic photometry using the code TA-DA \citep{dario2012tada} and assuming the BT-Settl synthetic spectra from \citet{allard2011}. We randomly displace each source in $T_{\rm eff}$ and $\log L_*$ from a uniform probability distribution in the range $-A_V/2$--$A_V/2$. We run simulations for two amounts of differential extinction: a low ($\Delta A_V=0.4$~mag) and an intermediate one ($\Delta A_V=1$~mag), and, as before, for two intrinsic age spreads: $\sigma\log t_{\rm true}$=0.2 and 0.4 dex. As for errors in $T_{\rm eff}$ (Section \ref{section:simulations-teff}), this uncertainty follows a type A bias (Section \ref{section:simulation_howtheywork}), since the relative excess $L_{\rm acc}/L_*$ is only weakly dependant on $A_V$. This holds in particular when $H\alpha$ excess together with optical fluxes are used to measure accretion \citep{demarchi2010}.

The results are shown in Figure \ref{fig:teffav_tautau}. Once again, the $\eta_{\rm meas}<\eta_{\rm true}$; however, the relative bias is smaller than that of simple $T_{\rm eff}$ errors (Figure \ref{fig:teff_tautau}). This is because the reddening vector, in $V-$ and $I-$ bands, is almost parallel to the low-mass PMS isochrones, and therefore extinction does not significantly alter the isochronal ages. This is not the case when restricting to intermediate-mass stars ($M>0.7$~M$_\odot$), for which we find $\eta_{\rm meas}$ is 20--30\% smaller than the $\eta_{\rm true}$.

\subsection{Evolutionary models}
Estimates of $\dot M_{\rm acc}$ and stellar parameters rely on stellar evolutionary models; hence, the derived $\eta$ remain model dependent.
There are open problems with current PMS isochrones, as the different families of models differ in their predictions \citep{hillenbrand2008} and often deviate systematically from the data \citep{mayne2007}.
In what follows we investigate if the adopted set of models, or an empirical calibration of them in the HRD, have important effects on the biases in $\eta$.

\subsubsection{Changing isochrones}
\label{appendix:evolmodels}
\begin{figure*}
\includegraphics[height=2\columnwidth,angle=90]{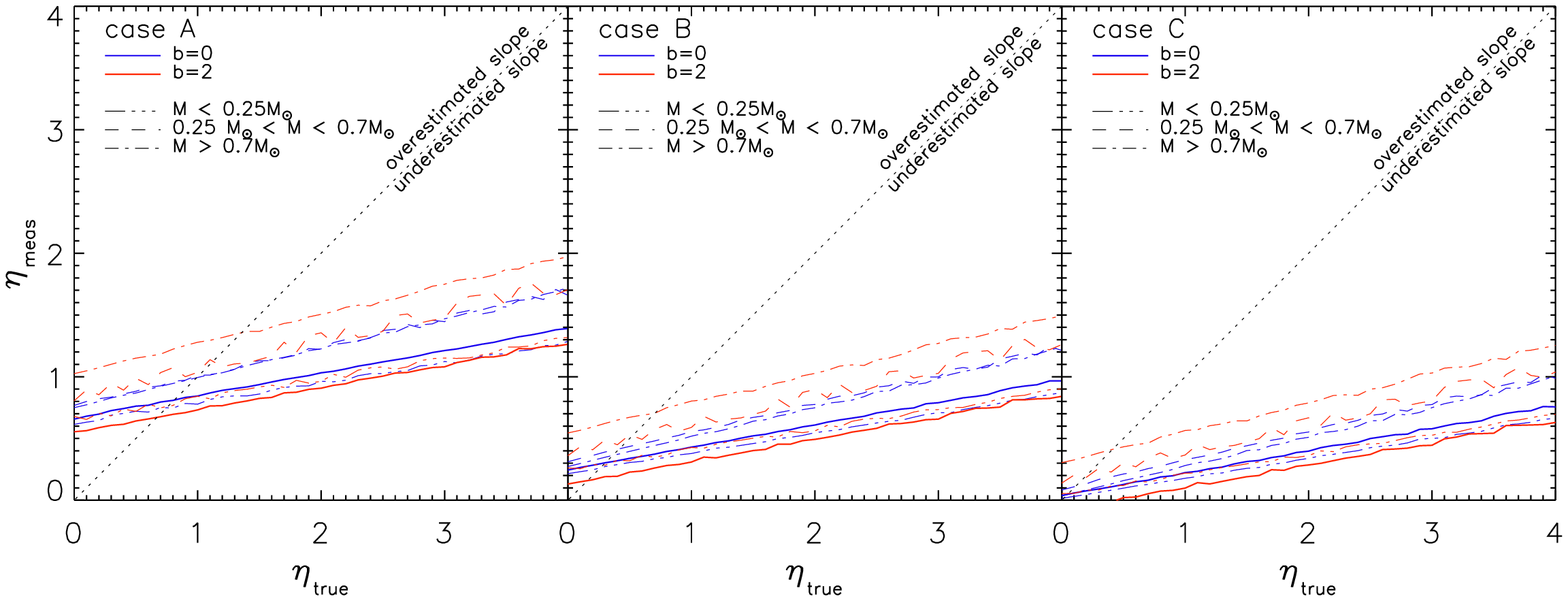}
\caption{Same as Figure \ref{fig:simplespread_tautau02}, but from a simulated population using \citet{dantona1998} isochrones and assuming a mean cluster age of 1~Myr.   \label{fig:simplespread_tautau02_dantona}}
\end{figure*}
 Throughout our simulations, we have used \citet{siess2000} isochrones; here we test if the bias in $\eta$ is heavily dependent on the assumed evolutionary models. To this end, we consider isochrones from \citet{dantona1998}, as they present some of the largest systematic differences compared to Siess. In particular, their models predict a much younger age for PMS stars, for instance, a mean age of 1~Myr for the ONC, less than half of that predicted by Siess isochrones. Thus, to generate a synthetic population in the HRD comparable with our previous simulations, we simulate cluster populations with a mean age of 1~Myr from \citet{dantona1998} isochrones.

For simplicity, we test only the case of a luminosity spread in the HRD, and use the same assumptions as in Appendix \ref{section:simulations-simplespread-ONC}, i.e., a real age spread $\sigma\log t=0.2$ and additional scatter in $\log L_*$ to reach an apparent measured $\sigma\log t=0.4$. The resulting $\eta_{\rm meas}$ versus $\eta_{\rm true}$ is shown in Figure \ref{fig:simplespread_tautau02_dantona}. The bias in $\eta$ for the 3 types of uncertainties is almost identical to that we obtained for the analogous simulations from Siess models (Figure \ref{fig:simplespread_tautau02}), despite the large systematic difference in the predicted absolute ages between these two grids of evolutionary models. This is due to the fact that the spacing in $\log L_*$ between contiguous isochrones (in $\log t$) remains similar even when different families of models show large offsets in their absolute age scales.

Our result shows that the ratio $\eta_{\rm true} / \eta_{\rm meas}$ is largely unaffected by the choice of evolutionary models. Furthermore, for a given cluster the value $\eta_{\rm meas}$ does not change much if one uses different evolutionary models; this was shown e.g., by \citet{manara2012} for the ONC, and also supported by the fact that measurements of $\eta$ in the literature, on a multitude of young clusters, and based on a heterogeneity of assumptions and models do not show noticeable differences in their results. Hence we argue that the true value of $\eta$ -- and the biases therein -- suggested by the modeling in this paper are not induced by our particular choice of adopting the \citet{siess2000} isochrones.

\subsubsection{Correction for systematic errors in the evolutionary models}
\label{section:acchistory}
\begin{figure}
\includegraphics[width=\columnwidth]{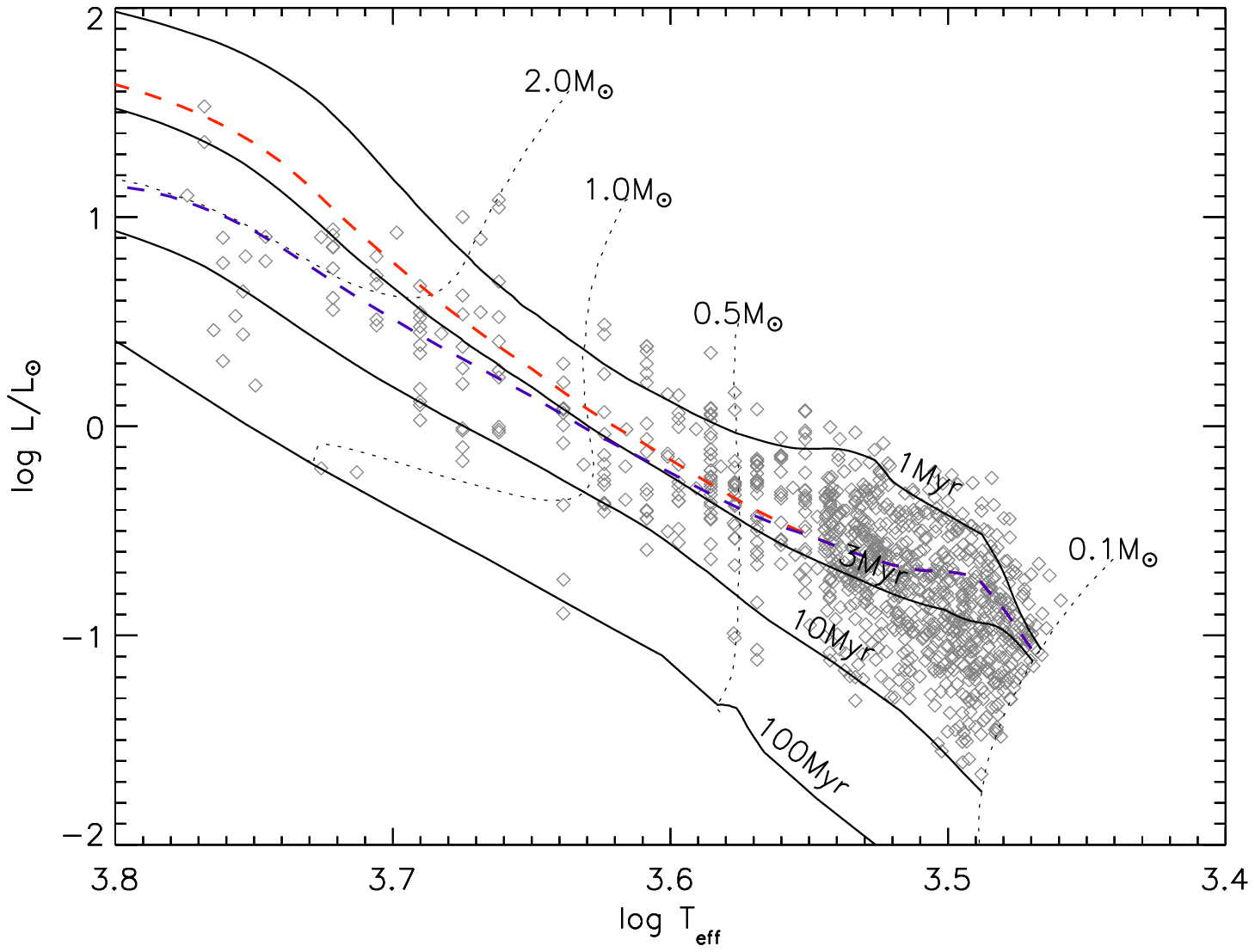}
\caption{HRD of the ONC from \citet{dario2010a}, with \citet{siess2000} evolutionary models. The dashed red line is the isochrone for the mean age of the cluster (2.5~Myr); the dashed blue line is the same after applying a $T_{\rm eff}$-dependent luminosity shift sufficient match the observed population.  \label{fig:acchistory_cmd}}
\end{figure}
\begin{figure*}
\includegraphics[width=1.6\columnwidth]{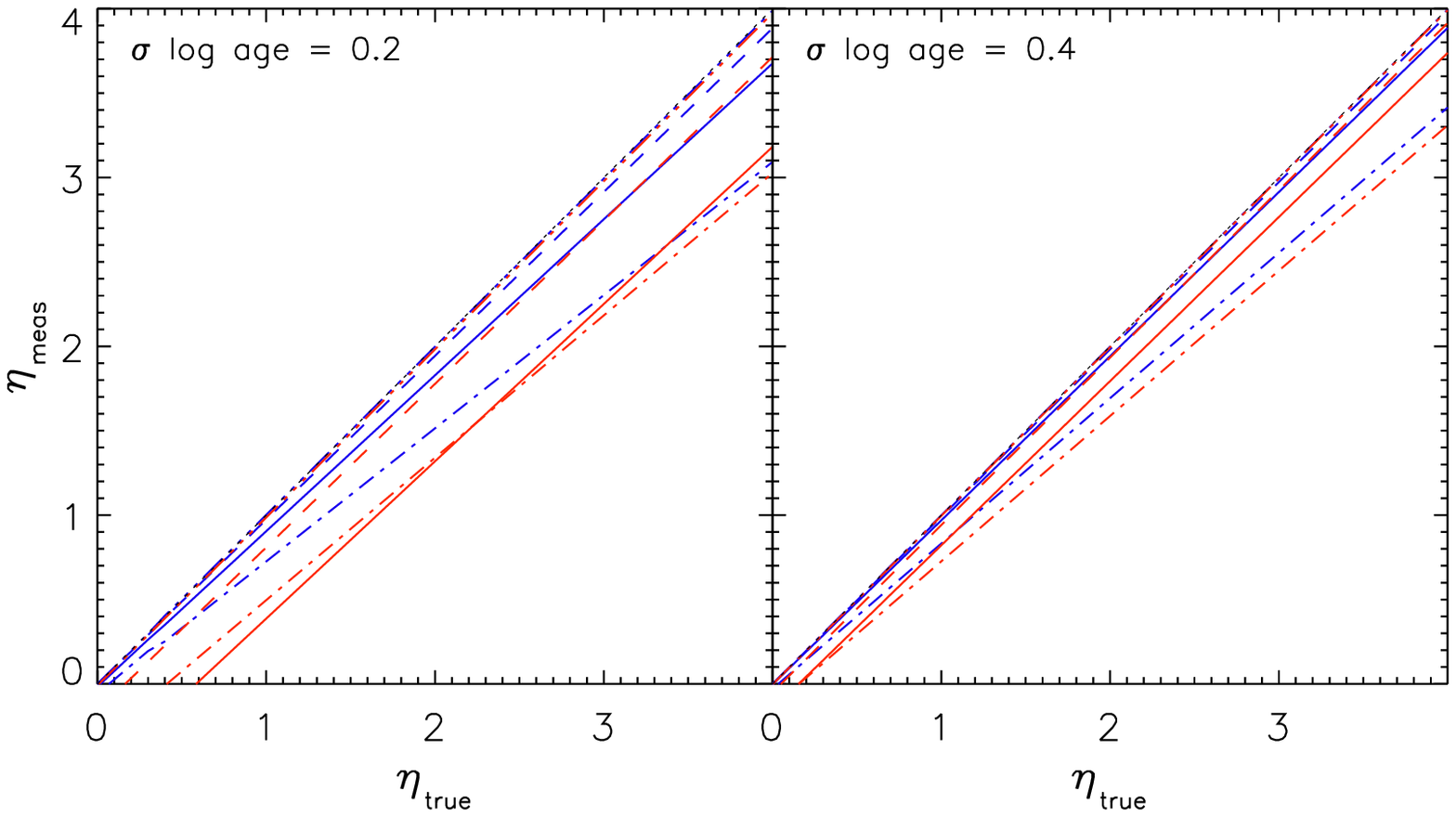}
\caption{Observed $\eta$ versus real $\eta$ when $L_*$ are shifted to fainter values at increasing masses to reproduce the age-mass dependence introduced by non-accreting isochrones, for 2 intrinsic age spreads. Colors and line styles are as in Figure \ref{fig:simplespread_tautau02}. \label{fig:tautau_acchistory}}
\end{figure*}

In the previous subsection we analyzed the effect of discrepancies between different evolutionary models. Here we focus on their general deviation from the observations. One of these discrepancies is the apparent correlation between mass and
average age for PMS clusters \citep{hartmann2003,dario2010b}, as in a
young system intermediate-mass stars appear generally older than
low-mass stars, by several Myr (see Figure \ref{fig:acchistory_cmd}.
%RDJ changed - check
If this systematic difference were real, it implies a significant
temporal delay of a few Myr in the formation of low-mass stars in a cluster. This
seems unlikely and the trend probably indicates an inaccurate prediction by the evolutionary
models.

There are many possible origins for this inconsistency. Several assumptions in the modeling of PMS mass tracks affect the predicted stellar parameters in different ways depending on the mass \citep{tognelli2011}: for example, reducing the mixing length could remove the mass-age correlation \citep{tognelli2013}. Alternatively, whereas standard PMS tracks are computed assuming an evolution without accretion, the inclusion of mass accretion in the modeling also affects the predicted parameters, in particular leads to smaller radii - therefore fainter luminosities \citep{tout1999,hosokawa2011}. This is expected to have a negligible effect for low-mass stars \citep{hosokawa2011}, but could be significant at masses $\geq 0.5\,M_{\odot}$.

We consider this second scenario, and assume that the standard isochrones overestimate the predicted $\log L_*$ for intermediate masses. We quantify the required correction empirically, based on the measured HRD of the ONC from \citet{dario2010a}. This is shown in Figure \ref{fig:acchistory_cmd}, together with \citet{siess2000} evolutionary models. The dashed red line is the isochrone corresponding to the mean isochronal age of the cluster (2.5~Myr). We have computed the required shift in $\log L_*$ from this isochrone, as a function of $T_{\rm eff}$, to match the mean observed luminosities; the resulting locus is the blue dashed line in Figure \ref{fig:acchistory_cmd}.
Thus, the two lines correspond, respectively, to the parameters that these models \emph{do} predict, and those they \emph{should} predict. Since observations, as well as our simulations in this paper, are based on the standard isochrones, this is a further bias we investigate.

For our purposes, it is practical to treat this systematic effect in reverse: we assume that the Siess isochrones are correct, and that some other effect shrinks the stellar radii of an amount equal to the difference between the two lines in Figure \ref{fig:acchistory_cmd}. We run a simulation as in the previous section, where in this case the positions of the simulated sources in the HRD are displaced with a rigid, $T_{\rm eff}$-dependent shift in $R_*$, instead of with a random scatter.
This case now follows a bias of type C (see Section \ref{section:simulation_howtheywork}, since effectively the measured stellar $R_*$ $\log L_*$ are correct, whereas $M_*$ and $t$ are wrongly inferred from the models.

The results are shown in Figure \ref{fig:tautau_acchistory}. Considering the entire sample (solid lines), $\eta_{\rm meas}$ is generally lower than $\eta_{\rm true}$. When we separate the mass ranges, it is evident that this is due mainly to the contribution of intermediate mass stars (dash-dotted lines), for which $\eta_{\rm true}$ is underestimated by 15--25\%.

\section{Selection effects}
\label{section:selection_effects}
\begin{figure*}
\includegraphics[width=2.\columnwidth]{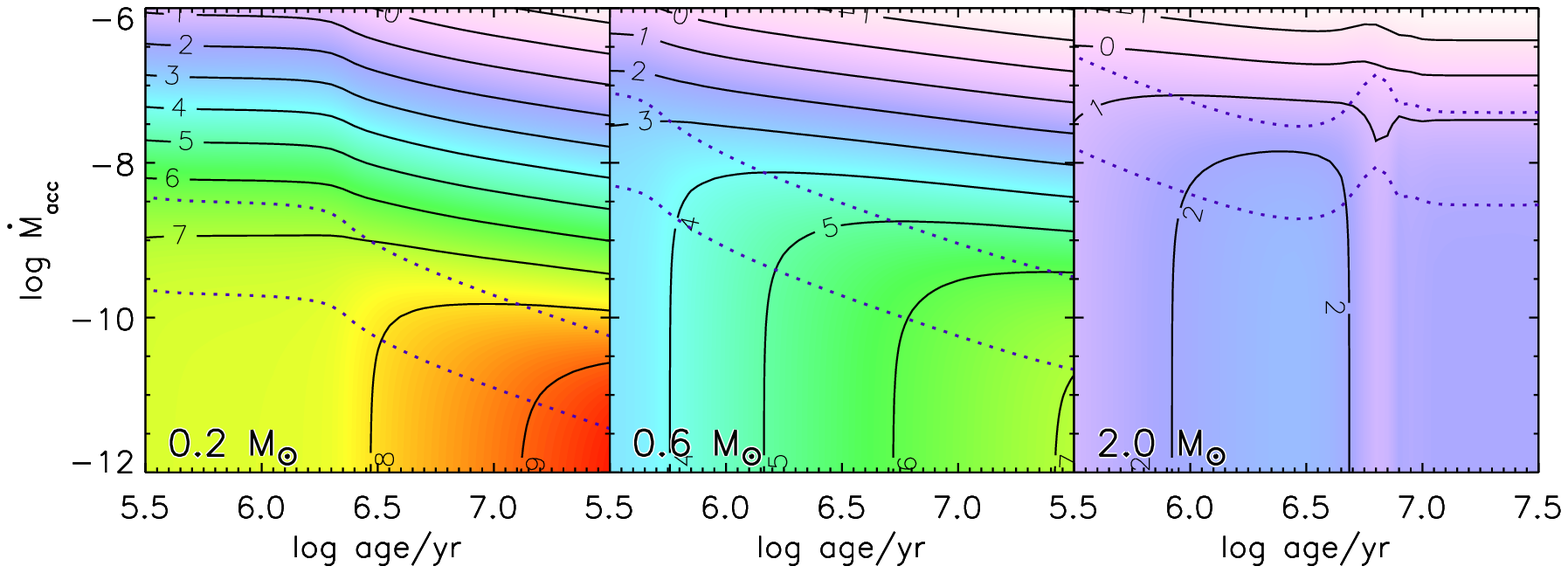}
\caption{Variation of the total observed absolute magnitude in the $H\alpha$ filter $F656N$, as a function for $\dot M_{\rm acc}$ and age, for three assumed masses ({\em solid contours}). The dotted lines indicate the loci where the measured $H\alpha$ excess is 0.1~mag (lower line) and 1~mag (upper line). \label{fig:selection_effects}}
\end{figure*}

\begin{figure*}
\includegraphics[width=1.6\columnwidth]{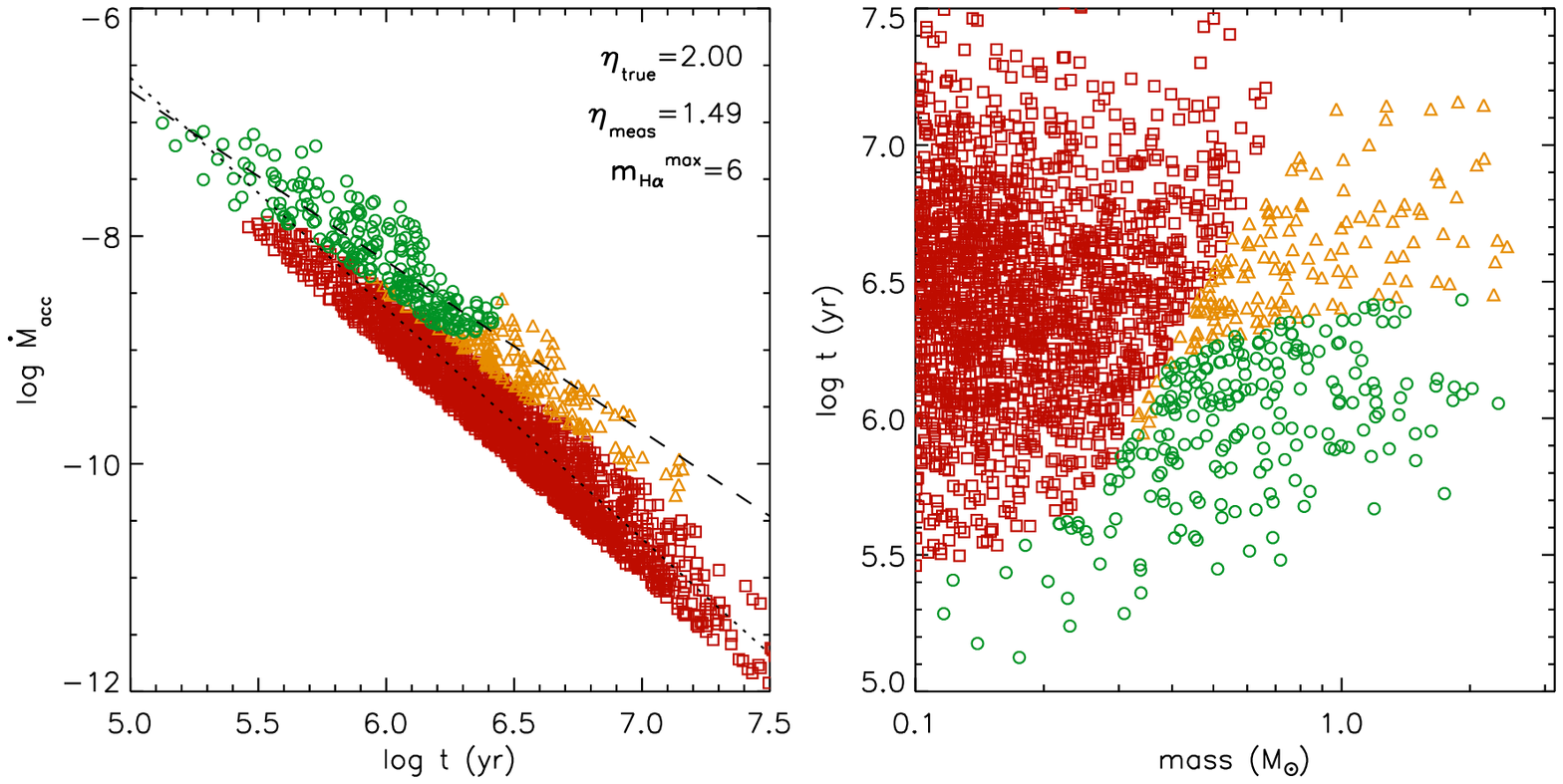}
\includegraphics[width=1.6\columnwidth]{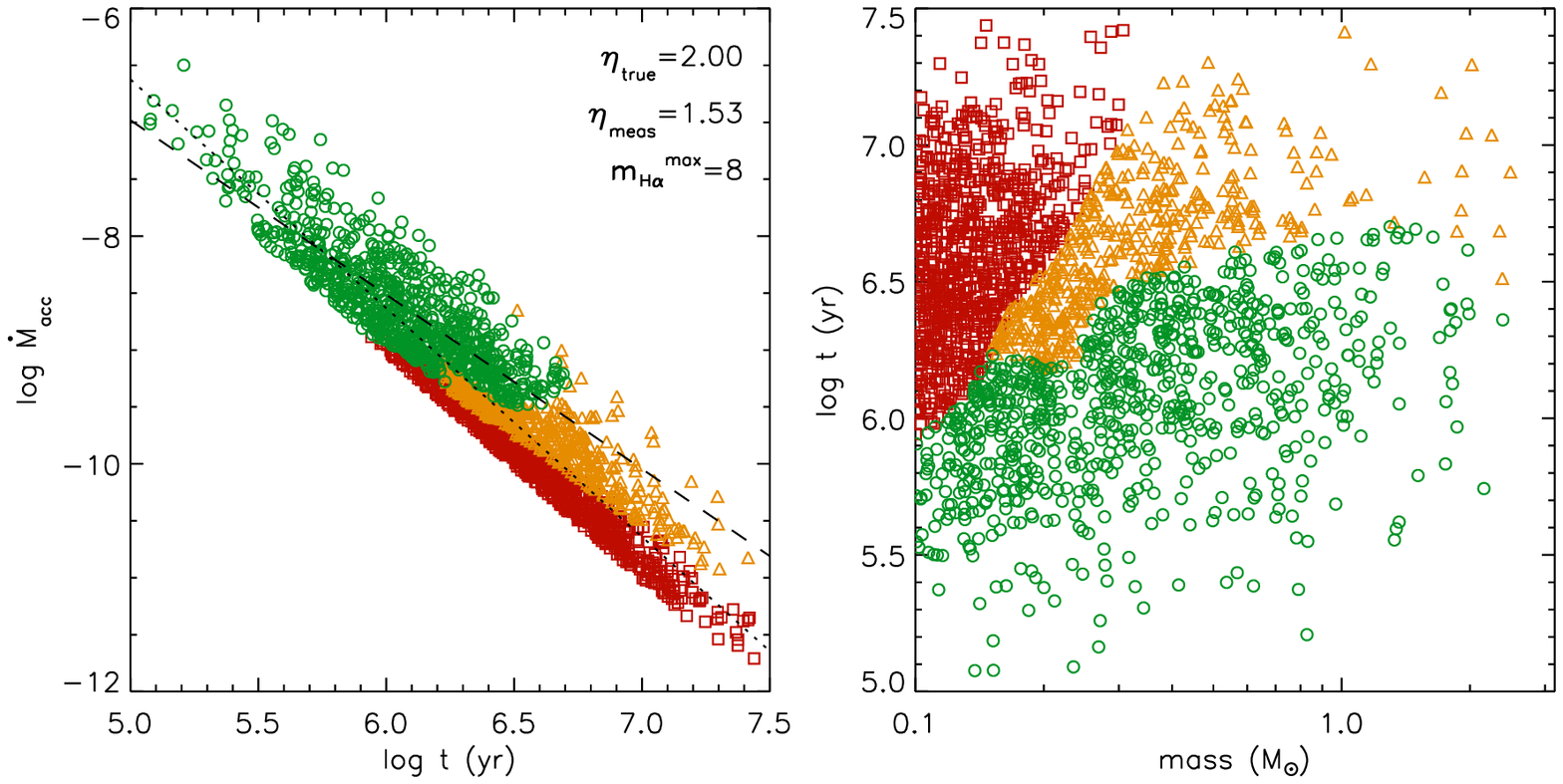}
\includegraphics[width=1.6\columnwidth]{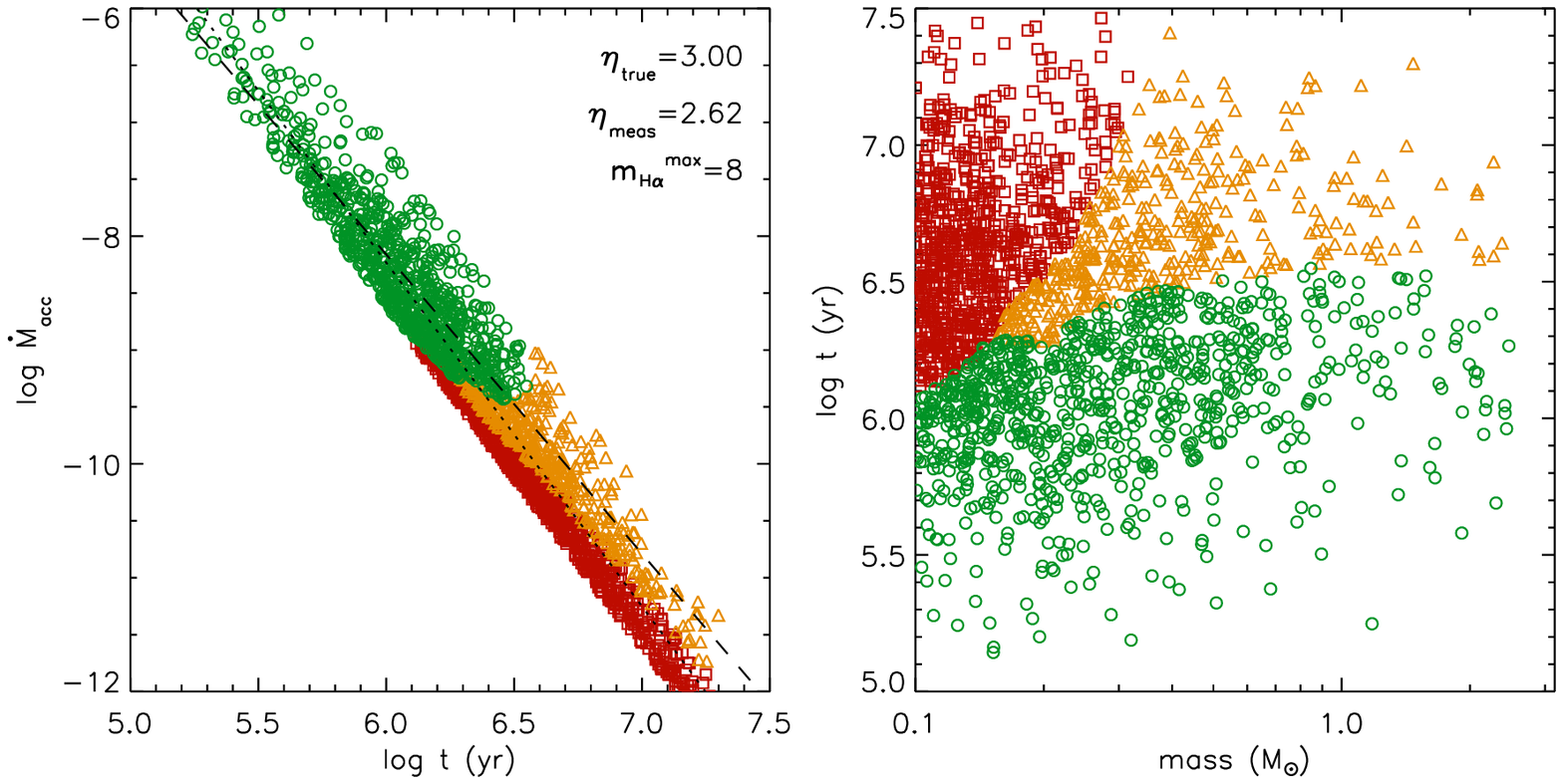}
\caption{Example of simulated selection effects introduced by $H\alpha$ detection, for a combination of 2 different detection limits in $F656N$: (m-M)=6~mag and 8~mag, and two input exponents $\eta_{\rm true}$ (dotted line). Red squares indicate fainter than the detection limit; orange triangles sources brighter than the detection limit, but with an excess $\Delta m_{{\rm H}\alpha}$ smaller than the photometric error; green circles the remaining detections with measurable excess. The dashed line is the fit of the latter, with a slope as indicated by $\eta_{\rm meas}$ in the legend. Sources correspond with each other between left and right panels.  \label{fig:selection_effects_population}}
\end{figure*}

\begin{figure*}
\includegraphics[width=1.6\columnwidth]{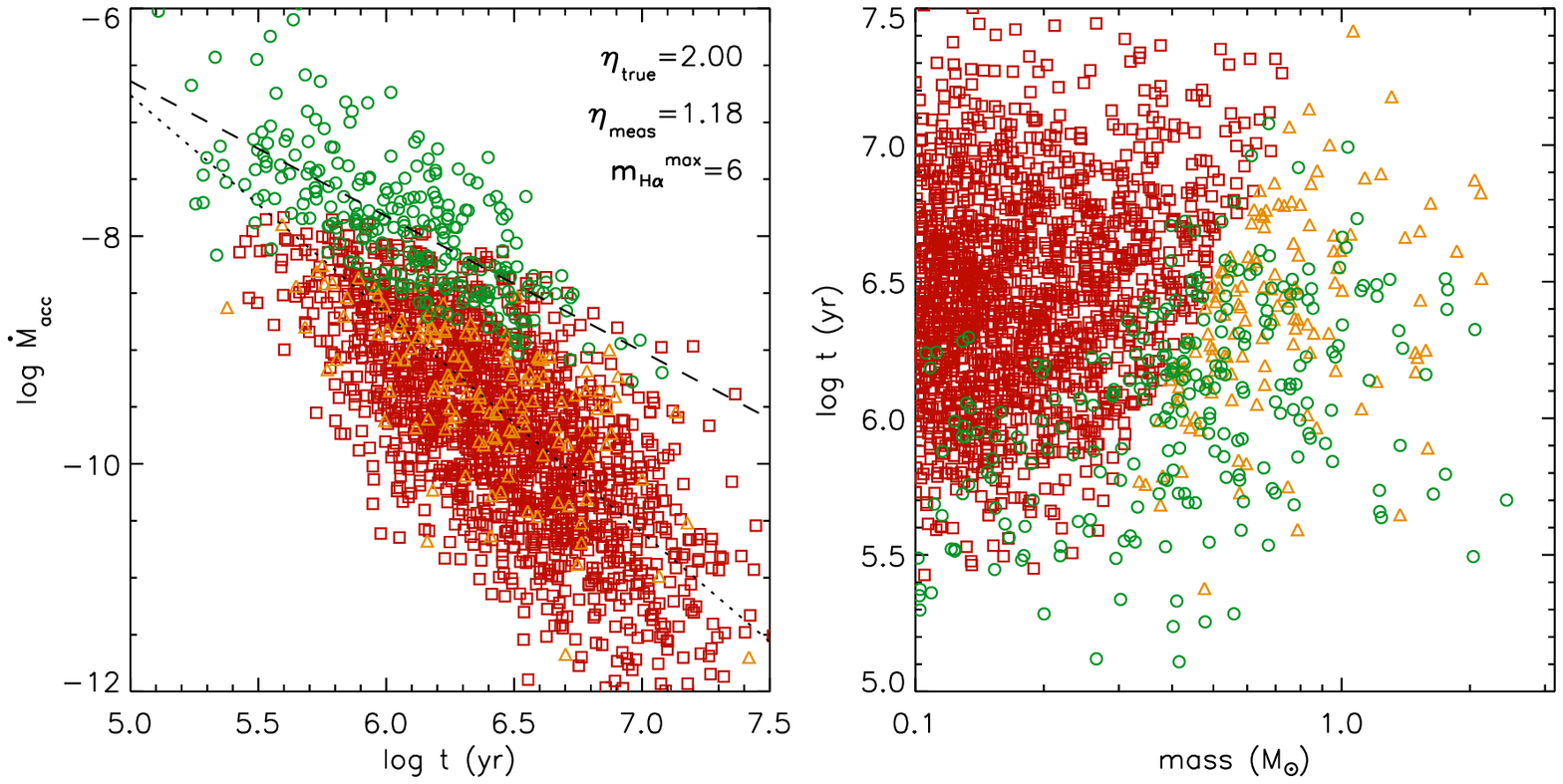}
\includegraphics[width=1.6\columnwidth]{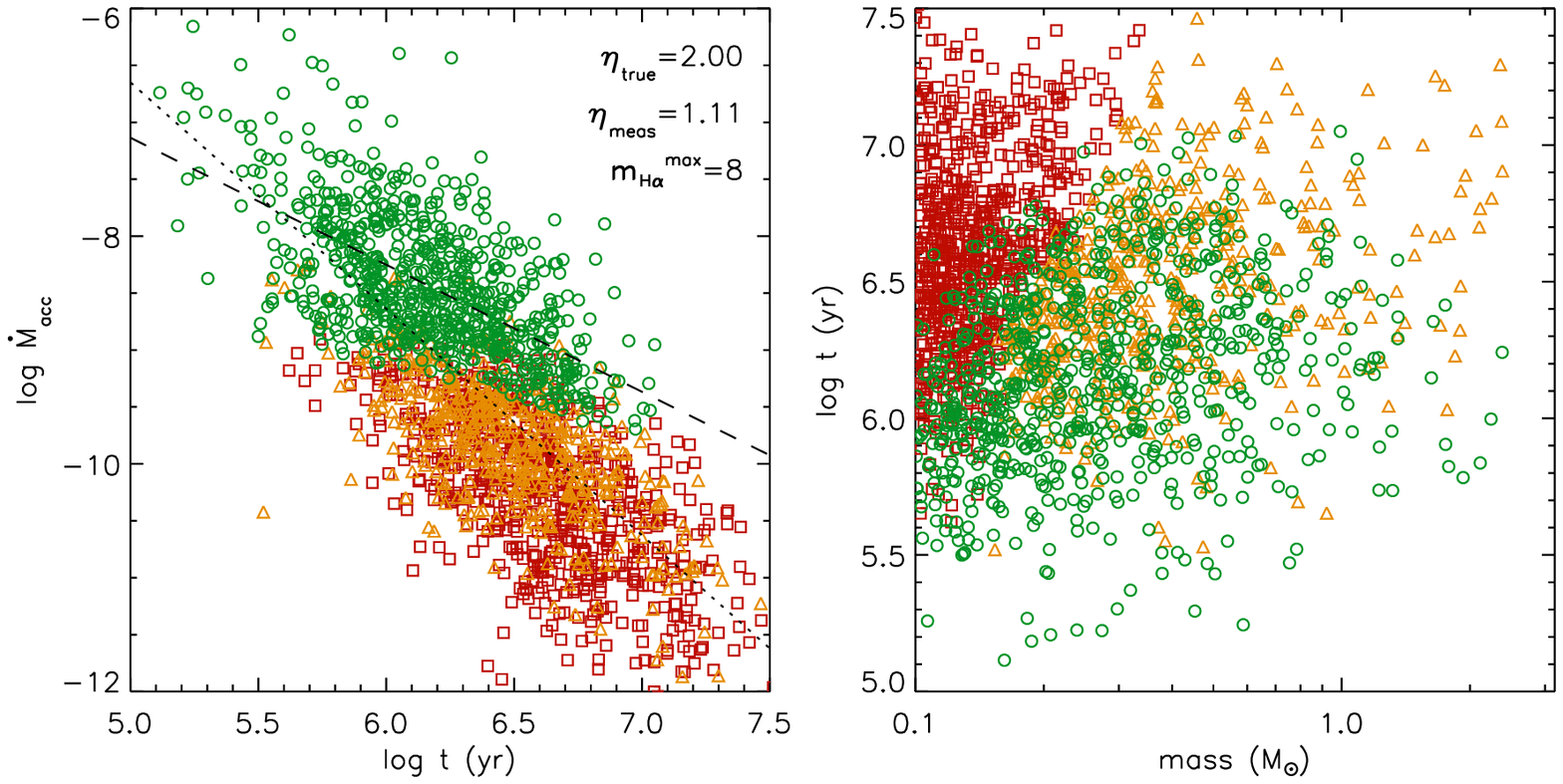}
\includegraphics[width=1.6\columnwidth]{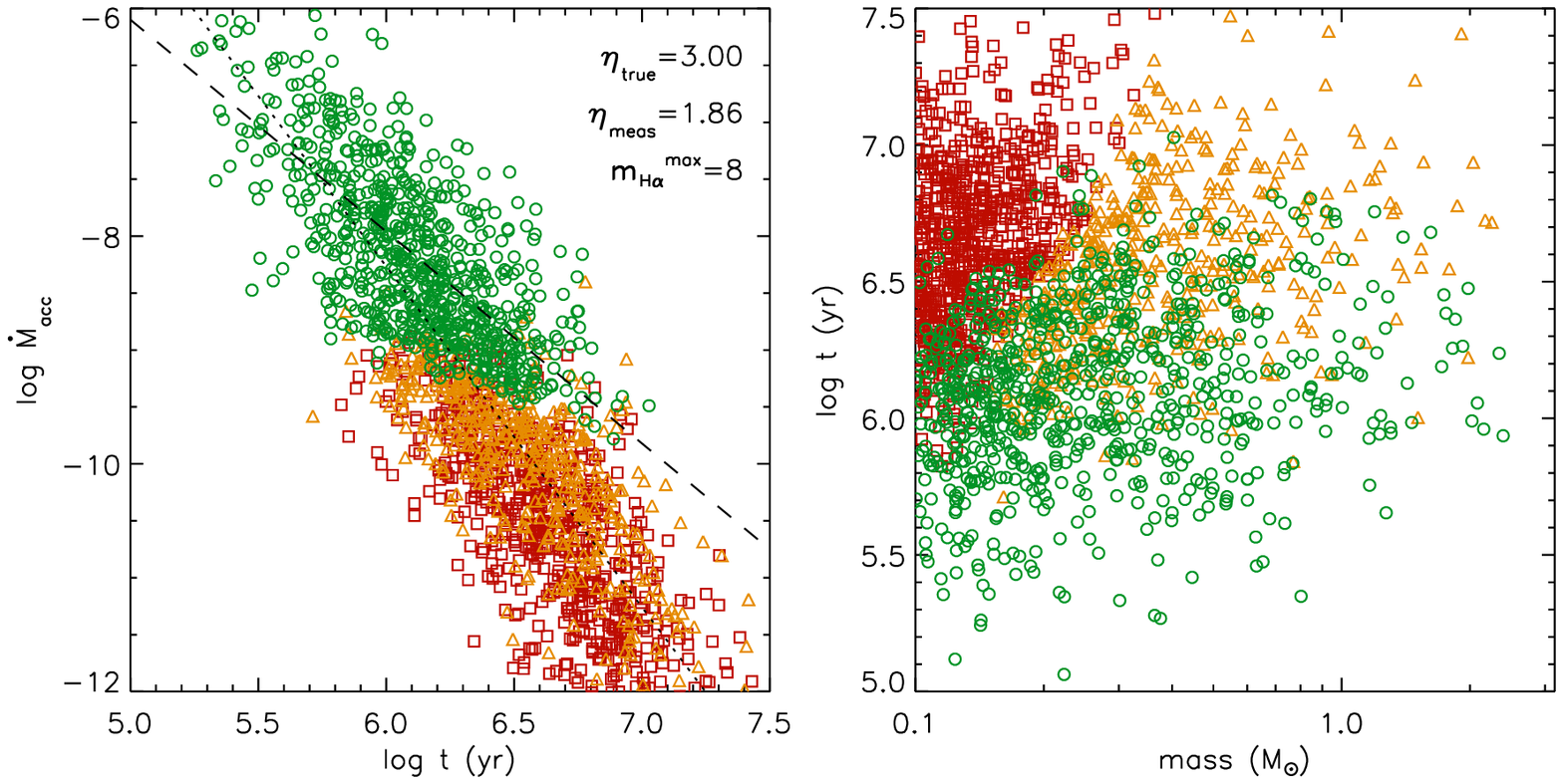}
\caption{Same as Figure \ref{fig:selection_effects_population}, for $b=1$ and after adding an additional random scatter to the accretion rates with $\sigma\log\dot M_{\rm accr}=0.65$~dex and constraining the photometric errors in H$\alpha$ to be $>0.1$~mag. \label{fig:selection_effects_population_scatter}}
\end{figure*}

As anticipated in Section \ref{section:intro}, sample incompleteness, from either lack of detection of stellar members, or lack of measurement of $\dot M_{\rm acc}$ for weak accretors, introduces selection effects in the $\dot M_{\rm acc}$--$t$ plane. Figure \ref{fig:selection_effects} shows a quantitative example of these effects. We have considered 3 stellar masses, 0.2, 0.6 and 2 M$_\odot$, and computed the variation of the stellar luminosity and accretion excess with age and $\dot M_{\rm acc}$. We assumed that the H$\alpha$ excess is the observational indicator to measure $\dot M_{\rm acc}$, and the specific technique is narrow band photometry with the $HST$ Wide Field Camera 3 filter $F656N$. We computed, for every $t$ and $\dot M_{\rm acc}$, both the measured excess in magnitudes in $F656N$ and the total absolute magnitude (photosphere and excess). This was done as follows: first, we applied equation \ref{equation:mdot2} to translate $\dot M_{\rm acc}$ into $L_{\rm acc}/L_*$. From the latter we derived $L_{{\rm H}\alpha}/L_*$ assuming the relation from \citet{demarchi2010}
\begin{equation}
\log L_{\rm acc}=\log L_{{\rm H}\alpha} + 1.72.
\end{equation}
\noindent Next, we derived the equivalent width of the $H\alpha$ line excess from:
\begin{equation}
EW_{{\rm H}\alpha}=\frac{L_{{\rm H}\alpha}}{L_*} \cdot \frac{\int_\lambda S_{\lambda}(T_{\rm eff}) d\lambda}{\int_{\lambda=\lambda(H\alpha)-0.5\AA}^{\lambda(H\alpha)+0.5\AA} S_{\lambda}(T_{\rm eff}) d\lambda}
\end{equation}
\noindent where $S_{\lambda}(T_{\rm eff})$ is the photospheric spectral energy distribution of the star, for which we assumed BT-Settl synthetic spectra from \citet{allard2011}. Last, we derived the associated (positive) excess in magnitudes from:
\begin{equation}
\Delta m_{{\rm H}\alpha}=2.5\cdot\log\bigg(1+\frac{EW_{{\rm H}\alpha}}{EW_{F656N}}\bigg)
\end{equation}
\noindent where $EW_{F656N}$ is the equivalent width of the filter.
We also computed, for every mass and age, the photospheric magnitude $m_{0,{\rm H}\alpha}$ by means of synthetic photometry using the tool TA-DA \citep{dario2012tada}, again using BT-Settl models.  Finally, the observed magnitude is the sum of photosphere and excess: $m_{{\rm H}\alpha}=m_{0,{\rm H}\alpha}-\Delta m_{{\rm H}\alpha}$.

The solid contours in Figure \ref{fig:selection_effects} indicate constant absolute magnitude: at very low $\dot M_{\rm acc}$ the excess is negligible, thus $m_{{\rm H}\alpha}$ is dominated by the photosphere and depends only on age, leading to vertical contours. At higher $\dot M_{\rm acc}$, however, the excess which correlates with the stellar parameters, leads to shallow correlations between $\dot M_{\rm acc}$ and age at a fixed magnitude. If the detection limit is shallow enough (for instance, $m=5$), in the low mass range (e.g. 0.2~M$_\odot$) the detection threshold will be inclined in $\dot M_{\rm acc}$ - $t$ plane, as the contours show. Depending on the depth of the sample, such a cut alters the underlying correlation between these two quantities.

The dotted contours in Figure \ref{fig:selection_effects}, on the other hand, indicate constant magnitude excess $\Delta m_{{\rm H}\alpha}$, respectively 0.1~mag (lower lines) and 1~mag (upper line). Again, these loci are inclined in the plane, so when the measurement of $\dot M_{\rm acc}$ is limited by the ability to isolate a small accretion-related excess from noises or other uncertainties, the sample of measured $\dot M_{\rm acc}$ will suffer from a correlated selection effect with age. As shown, in this case the problem dominates at large masses, where high values of $\dot M_{\rm acc}$ produce small flux excesses.

We model a more realistic representation of these effects to understand their overall influence on $\eta_{\rm meas}$ for a stellar population. We simulate a young population as in Section \ref{section:simulations}, assuming a mean age $t=2.5$~Myr, an age spread $\sigma \log t_{\rm true}=0.4$~dex, and assign values of $\dot M_{\rm acc}$ assuming with b=2 and variable $\eta_{\rm true}$. For each generated star, considering its stellar parameters and $\dot M_{\rm acc}$, we compute both the magnitude excess in $H\alpha$, $\Delta m_{{\rm H}\alpha}$, and the total (absolute) magnitude $m_{{\rm H}\alpha}$ as we did before for Figure \ref{fig:selection_effects}.
We introduce a detection limit $m_{{\rm H}\alpha}^{max}$, for which the assumed signal to noise ratio is 3 and the completeness abruptly drops from 100\% to 0. This simplified choice allows us to isolate the general bias due selection effects without adding unnecessary scatter due to the stochasticity of source detection. We assign to every magnitude $m_{{\rm H}\alpha}$ a photometric error, assuming that it is only due to shot noise from photon counting, thus:
\begin{equation}
\label{equation:photerrHa}
\sigma(m_{{\rm H}\alpha})=2.5\log\bigg(1+\frac{10^{-0.2(m_{{\rm H}\alpha}^{max} - m_{{\rm H}\alpha})}}{3}\bigg)
\end{equation}
As another general assumption, we assume that H$\alpha$ photometry is responsible for all the incompleteness and photometric uncertainty affecting the sample. Since in practice photometric estimates of $\dot M_{\rm acc}$ generally require additional broad band photometry to isolate the excess (e.g., $V$ and $I$, \citealt{demarchi2010}), this implies the latter should be significantly deeper than H$\alpha$.

Figure \ref{fig:selection_effects_population} shows the result for 3 combinations of $\eta_{\rm true}$ and $m_{{\rm H}\alpha}^{max}$ producing a significant incompleteness in the PMS mass range. Depending on the magnitude limit, some stars are too faint to be detected. Also, another fraction are detected, but their accretion excess is smaller than the photometric error, leading to an average estimate of $\dot M_{\rm acc}$ compatible with 0 and the removal of these sources from the $\log\dot M_{\rm acc}$--$\log t$ plane. The remaining detections show a correlation of $\dot M_{\rm acc}$ with age, determined through linear fit, which differs from the real one, and is much shallower for the 3 cases shown in Figure \ref{fig:selection_effects_population}. We have inspected a broad range of assumed $\eta_{\rm true}$, $b$ and $m_{{\rm H}\alpha}^{max}$; the examples in Figure \ref{fig:selection_effects_population} appear representative of the general behavior, except if $0\sim\eta_{\rm true}\lesssim1$, in which case the diagonal threshold in the $\dot M_{\rm acc}$ increases $\eta_{\rm meas}$ by a few tenths.

The scatter of the $\dot M_{\rm acc}$ values at a given age shown in Figure \ref{fig:selection_effects_population} is solely due to the mass dependence introduced by $b=2$. In reality an additional scatter at any $M_*$ and $t$ is expected, as the measured $\dot M_{\rm acc}$ depends not only on $M_*$ and $t$, but can also vary star by star due to differences in the disk properties, temporal variability of $\dot M_{\rm acc}$, or due to poorly understood correlations with the stellar parameters (see e.g., Figure \ref{fig:simplespread_mdot_age}). We test if this modifies the bias due to selection effects. We repeat the same simulation adding a random gaussian scatter to $\log \dot M_{\rm acc}$ before computing the photospheric and excess magnitudes. We assume a $1\sigma$ width of 0.65 dex: this is the extent of the random star-to-star variations of $\dot M_{\rm acc}$ needed to obtain a total standard deviation of the data points in the $\log\dot M_{\rm acc}$--$\log t$ plane that matches the 0.85 dex found in the ONC by \citet{manara2012}. We also add 0.1~mag in quadrature to the assumed photometric errors $\sigma(m_{{\rm H}\alpha})$ from Equation \ref{equation:photerrHa}. This is to avoid the results being  altered by unrealistically small photometric errors at bright magnitudes.
The results are shown in Figure \ref{fig:selection_effects_population_scatter}; data corresponding to good detections and non detections appear scattered, but the general trends are not very different from those of Figure \ref{fig:selection_effects_population}.

We cannot derive a general correction for these types of selection effects, as in reality they depend on multiple factors: e.g., the detection completeness functions and the photometric error distribution in each band, as well as the other overall uncertainties affecting both magnitudes and stellar parameters, and the possible inaccuracy of the atmosphere models we assumed in our modeling. However, if the studied stellar population is characterized to a much fainter luminosity than those for which $\dot M_{\rm}$ is assessable (e.g., through broad band photometry), it is useful to study the fraction of accretors as a function of age (and mass). This is because, if the age spread within a region is sufficiently small, one would not expect any significant variation of the fraction of detected accretors with age or mass. If on the contrary, as evident from the right panels of Figures \ref{fig:selection_effects_population} and \ref{fig:selection_effects_population_scatter}, at low masses the detected accretors tend to be all younger than the cluster average age, strong selection effects in the samples may be present and bias $\eta_{\rm meas}$.


\begin{thebibliography}{}
\bibitem[Alcal{\'a} et al.(2014)]{alcala2013} Alcal{\'a}, J.~M., Natta, A., Manara, C.~F., et al.\ 2014, A\&A, 561, A2
\bibitem[Alencar et al.(2005)]{alencar2005} Alencar, S.~H.~P., Basri, G., Hartmann, L., \& Calvet, N.\ 2005, A\&A, 440, 595
\bibitem[Allard et al.(2011)]{allard2011} Allard, F., Homeier, D., \& Freytag, B.\ 2011, 16th Cambridge Workshop on Cool Stars, Stellar Systems, and the Sun, 448, 91
\bibitem[Antoniucci et al.(2011)]{antoniucci2011} Antoniucci, S., Garc{\'{\i}}a L{\'o}pez, R., Nisini, B., et al.\ 2011, A\&A, 534, A32
\bibitem[Baraffe et al.(2012)]{baraffe2012} Baraffe, I., Vorobyov, E., \& Chabrier, G.\ 2012, ApJ, 756, 118
\bibitem[Baraffe et al.(2009)]{baraffe2009} Baraffe, I., Chabrier, G., \& Gallardo, J.\ 2009, ApJL, 702, L27
\bibitem[Bastian et al.(2010)]{bastian2010} Bastian, N., Covey, K.~R., \& Meyer, M.~R.\ 2010, ARA\&A, 48, 339
\bibitem[Beccari et al.(2010)]{beccari2010} Beccari, G., Spezzi, L., De Marchi, G., et al.\ 2010, ApJ, 720, 1108
\bibitem[Bell et al.(2013)]{bell2013} Bell, C.~P.~M., Naylor, T., Mayne, N.~J., Jeffries, R.~D., \& Littlefair, S.~P.\ 2013, MNRAS, 434, 806
\bibitem[Biazzo et al.(2012)]{biazzo2012} Biazzo, K., Alcal{\'a}, J.~M., Covino, E., et al.\ 2012, A\&A, 547, A104
\bibitem[Biazzo et al.(2009)]{biazzo2009} Biazzo, K., Melo, C.~H.~F., Pasquini, L., et al.\ 2009, A\&A, 508, 1301
\bibitem[Bouvier et al.(2003)]{bouvier2003} Bouvier, J., Grankin, K.~N., Alencar, S.~H.~P., et al.\ 2003, A\&A, 409, 169
\bibitem[Calvet \& Gullbring(1998)]{calvet-gullbring1998} Calvet, N., \& Gullbring, E.\ 1998, ApJ, 509, 802
\bibitem[Calvet et al.(2004)]{calvet2004} Calvet, N., Muzerolle, J., Brice{\~n}o, C., et al.\ 2004, AJ, 128, 1294
\bibitem[Cardelli et al.(1989)]{cardelli1989} Cardelli, J.~A., Clayton, G.~C., \& Mathis, J.~S.\ 1989, ApJ, 345, 245
\bibitem[Clarke \& Pringle(2006)]{clarke2006} Clarke, C.~J., \& Pringle, J.~E.\ 2006, MNRAS, 370, L10
\bibitem[Costigan et al.(2012)]{costigan2012} Costigan, G., Scholz, A., Stelzer, B., et al.\ 2012, MNRAS, 427, 1344
\bibitem[D'Antona \& Mazzitelli (1998)]{dantona1998} D’Antona F., Mazzitelli I., 1998, in Rebolo R., Martin E. L., Zapatero Osorio M. R., eds, ASP Conf. Ser. 134: Brown Dwarfs and Extrasolar Planets A Role for Superadiabatic Convection in Low Mass Structures. pp 442
\bibitem[Da Rio \& Robberto(2012)]{dario2012tada} Da Rio, N., \& Robberto, M.\ 2012, AJ, 144, 176
\bibitem[Da Rio et al.(2012)]{dario2012} Da Rio, N., Robberto, M., Hillenbrand, L.~A., Henning, T., \& Stassun, K.~G.\ 2012, ApJ, 748, 14
\bibitem[Da Rio et al.(2010b)]{dario2010b} Da Rio, N., Gouliermis, D.~A., \& Gennaro, M.\ 2010, ApJ, 723, 166
\bibitem[Da Rio et al.(2010a)]{dario2010a} Da Rio, N., Robberto, M., Soderblom, D.~R., et al.\ 2010, ApJ, 722, 1092
\bibitem[Da Rio et al.(2009)]{dario2009} Da Rio, N., Robberto, M., Soderblom, D.~R., et al.\ 2009, ApJS, 183, 261
\bibitem[Dahm(2008)]{dahm2008} Dahm, S.~E.\ 2008, AJ, 136, 521
\bibitem[de Juan Ovelar et al.(2012)]{dejuanovelar2012} de Juan Ovelar, M., Kruijssen, J.~M.~D., Bressert, E., et al.\ 2012, A\&A, 546, L1
\bibitem[De Marchi et al.(2012)]{demarchi2012} De Marchi, G., Panagia, N., Guarcello, M.~G., \& Bonito, R.\ 2012, American Astronomical Society Meeting Abstracts \#219, 2 \#337.06
\bibitem[De Marchi et al.(2011b)]{demarchi2011b} De Marchi, G., Panagia, N., Romaniello, M., et al.\ 2011, ApJ, 740, 11
\bibitem[De Marchi et al.(2011a)]{demarchi2011a} De Marchi, G., Paresce, F., Panagia, N., et al.\ 2011, ApJ, 739, 27
\bibitem[De Marchi et al.(2010)]{demarchi2010} De Marchi, G., Panagia, N., \& Romaniello, M.\ 2010, ApJ, 715, 1
\bibitem[Edwards et al.(2006)]{edwards2006} Edwards, S., Fischer, W., Hillenbrand, L., \& Kwan, J.\ 2006, ApJ, 646, 319
\bibitem[Ercolano et al.(2013)]{ercolano2013} Ercolano, B., Mayr, D., Owen, J.~E., Rosotti, G., \& Manara, C.~F.\ 2013, MNRAS, {\em in press} (arXiv:1312.3154)
\bibitem[Fang et al.(2013)]{fang2013} Fang, M., Kim, J.~S., van Boekel, R., et al.\ 2013, ApJS, 207, 5
\bibitem[Fang et al.(2009)]{fang2009} Fang, M., van Boekel, R., Wang, W., et al.\ 2009, A\&A, 504, 461
\bibitem[Fedele et al.(2010)]{fedele2010} Fedele, D., van den Ancker, M.~E., Henning, T., Jayawardhana, R., \& Oliveira, J.~M.\ 2010, A\&A, 510, A72
\bibitem[Fischer et al.(2011)]{fischer2011} Fischer, W., Edwards, S., Hillenbrand, L., \& Kwan, J.\ 2011, ApJ, 730, 73
\bibitem[Frost \& Thompson (2000)]{frost2000} Frost, C., \& Thompson, S.~G.\ 2000, Journal of the Royal Statistical Society, Series A, 163, 173-18
\bibitem[Garcia Lopez et al.(2006)]{garcialopez2006} Garcia Lopez, R., Natta, A., Testi, L., \& Habart, E.\ 2006, A\&A, 459, 837
\bibitem[Gatti et al.(2008)]{gatti2008} Gatti, T., Natta, A., Randich, S., Testi, L., \& Sacco, G.\ 2008, A\&A, 481, 423
\bibitem[Guarcello et al.(2010)]{guarcello2010} Guarcello, M.~G., Damiani, F., Micela, G., et al.\ 2010, A\&A, 521, A18
\bibitem[Gullbring et al.(1998)]{gullbring1998} Gullbring, E., Hartmann, L., Briceno, C., \& Calvet, N.\ 1998, ApJ, 492, 323
\bibitem[Gullbring et  al.(1997)]{gullbring1997} Gullbring, E., Barwig, H., \& Schmitt, J.~H.~M.~M.\ 1997, A\&A, 324, 155
\bibitem[Haisch et al.(2001)]{haisch2001} Haisch, K.~E., Jr., Lada, E.~A., \& Lada, C.~J.\ 2001, ApJL, 553, L153
\bibitem[Hartmann et al.(1997)]{hartmann1997} Hartmann, L., Cassen, P., \& Kenyon, S.~J.\ 1997, ApJ, 475, 770
\bibitem[Hartmann et al.(1998)]{hartmann1998} Hartmann, L., Calvet, N., Gullbring, E., \& D'Alessio, P.\ 1998, ApJ, 495, 385
\bibitem[Hartmann(2001)]{hartmann2001} Hartmann, L.\ 2001, AJ, 121, 1030
\bibitem[Hartmann(2003)]{hartmann2003}Hartmann, L. 2003, ApJ, 585, 398
\bibitem[Hartmann et al.(2011)]{hartmann2011} Hartmann, L., Zhu, Z., \& Calvet, N.\ 2011, arXiv:1106.3343
\bibitem[Herbst et al.(2002)]{herbst2002} Herbst, W., Bailer-Jones, C.~A.~L., Mundt, R., Meisenheimer, K., \& Wackermann, R.\ 2002, A\&A, 396, 513
\bibitem[Herczeg \& Hillenbrand(2008)]{herczeg-hillenbrand2008} Herczeg, G.~J., \& Hillenbrand, L.~A.\ 2008, ApJ, 681, 594
\bibitem[Hern{\'a}ndez et al.(2008)]{hernandez2008} Hern{\'a}ndez, J., Hartmann, L., Calvet, N., et al.\ 2008, ApJ, 686, 1195
\bibitem[Hillenbrand et al.(2008)]{hillenbrand2008} Hillenbrand, L.~A., Bauermeister, A., \& White, R.~J.\ 2008, 14th Cambridge Workshop on Cool Stars, Stellar Systems, and the Sun, 384, 200
\bibitem[Hillenbrand \& Hartmann(1998)]{hillenbrand1998} Hillenbrand, L.~A., \& Hartmann, L.~W.\ 1998, ApJ, 492, 540
\bibitem[Hillenbrand(1997)]{hillenbrand1997} Hillenbrand, L.~A.\ 1997, AJ, 113, 1733
\bibitem[Hosokawa et al.(2011)]{hosokawa2011} Hosokawa, T., Offner, S.~S.~R., \& Krumholz, M.~R.\ 2011, ApJ, 738, 140
\bibitem[Ingleby et al.(2013)]{ingleby2013} Ingleby, L., Calvet, N., Herczeg, G., et al.\ 2013, ApJ, 767, 112
\bibitem[Jayawardhana et al.(2006)]{jayawardhana2006} Jayawardhana, R., Coffey, J., Scholz, A., Brandeker, A., \& van Kerkwijk, M.~H.\ 2006, ApJ, 648, 1206
\bibitem[Jeffries et al.(2011)]{jeffries2011} Jeffries, R.~D., Littlefair, S.~P., Naylor, T., \& Mayne, N.~J.\ 2011, MNRAS, 418, 1948
\bibitem[Jeffries(2007)]{jeffries2007} Jeffries, R.~D.\ 2007, MNRAS, 381, 1169
\bibitem[Koenigl(1991)]{koenigl1991} Koenigl A., 1991, ApJL, 370, L39
\bibitem[Koepferl et al.(2012)]{koepferl2012} Koepferl, C.~M., Ercolano, B., Dale, J., et al.\ 2012, MNRAS, 213
\bibitem[Kraus \& Hillenbrand(2009)]{kraus-hillenbrand2009} Kraus, A. L., \& Hillenbrand, L. A. 2009, ApJ, 704, 531
\bibitem[Kroupa(2001)]{kroupa2001} Kroupa, P.\ 2001, MNRAS, 322, 231
\bibitem[Krumholz \& Tan(2007)]{krumholz2007} Krumholz, M.~R., \& Tan, J.~C.\ 2007, ApJ, 654, 304
\bibitem[Kudryavtseva et al.(2012)]{kudryavtseva2012} Kudryavtseva, N., Brandner, W., Gennaro, M., et al.\ 2012, ApJL, 750, L44
\bibitem[Kurosawa et al.(2008)]{kurosawa2008} Kurosawa, R., Romanova, M.~M., \& Harries, T.~J.\ 2008, MNRAS, 385, 1931
\bibitem[Littlefair et al.(2011)]{littlefair2011} Littlefair, S.~P., Naylor, T., Mayne, N.~J., Saunders, E., \& Jeffries, R.~D.\ 2011, MNRAS, 413, L56
\bibitem[Luhman(1999)]{luhman1999} Luhman, K.~L.\ 1999, ApJ, 525, 466
\bibitem[Manara et al.(2013)]{manara2013} Manara, C.~F., Beccari, G., Da Rio, N., et al.\ 2013, A\&A, \emph{in press} (arXiv:1307.8118)
\bibitem[Manara et al.(2012)]{manara2012} Manara, C.~F., Robberto, M., Da Rio, N., et al.\ 2012, ApJ, 755, 154
\bibitem[Mayne \& Harries(2010)]{mayne-harries2010} Mayne, N.~J., \& Harries, T.~J.\ 2010, MNRAS, 409, 1307
\bibitem[Mayne et al.(2007)]{mayne2007} Mayne, N.~J., Naylor, T., Littlefair, S.~P., Saunders, E.~S., \& Jeffries, R.~D.\ 2007, MNRAS, 375, 1220
\bibitem[Mohanty et al.(2005)]{mohanty2005} Mohanty, S., Jayawardhana, R., \& Basri, G.\ 2005, ApJ, 626, 498
\bibitem[Muzerolle et al.(2005)]{muzerolle2005} Muzerolle, J., Luhman, K.~L., Brice{\~n}o, C., Hartmann, L., \& Calvet, N.\ 2005, ApJ, 625, 906
\bibitem[Muzerolle et al.(2003)]{muzerolle2003} Muzerolle, J., Hillenbrand, L., Calvet, N., Brice{\~n}o, C., \& Hartmann, L.\ 2003, ApJ, 592, 266
\bibitem[Natta et al.(2006)]{natta2006} Natta, A., Testi, L., \& Randich, S.\ 2006, A\&A, 452, 245
\bibitem[Naylor(2009)]{naylor2009} Naylor, T.\ 2009, MNRAS, 399, 432
\bibitem[Nguyen et al.(2009)]{nguyen2009} Nguyen, D.~C., Scholz, A., van Kerkwijk, M.~H., Jayawardhana, R., \& Brandeker, A.\ 2009, ApJL, 694, L153
\bibitem[Owen et al.(2013)]{owen2013} Owen, J.~E., Hudoba de Badyn, M., Clarke, C.~J., \& Robins, L.\ 2013, MNRAS, 436, 1430
\bibitem[Parmentier \& Pfalzner(2013)]{parmentier2013} Parmentier, G., \& Pfalzner, S.\ 2013, A\&A, 549, A132
\bibitem[Pascucci et al.(2012)]{pascucci2012} Pascucci, I., Gorti, U., \& Hollenbach, D.\ 2012, ApJL, 751, L42
\bibitem[Pascucci \& Sterzik(2009)]{pascucci2009} Pascucci, I., \& Sterzik, M.\ 2009, ApJ, 702, 724
\bibitem[Preibisch(2012)]{preibish2012} Preibisch, T.\ 2012, Research in Astronomy and Astrophysics, 12, 1
\bibitem[Rebull et al.(2006)]{rebull2006} Rebull, L.~M., Stauffer, J.~R., Megeath, S.~T., Hora, J.~L., \& Hartmann, L.\ 2006, ApJ, 646, 297
\bibitem[Reggiani et al.(2011)]{reggiani2011} Reggiani, M., Robberto, M., Da Rio, N., et al.\ 2011, A\&A, 534, A83
\bibitem[Rigliaco et al.(2012)]{rigliaco2012} Rigliaco, E., Natta, A., Testi, L., et al.\ 2012, A\&A, 548, A56
\bibitem[Rigliaco et al.(2011)]{rigliaco2011} Rigliaco, E., Natta, A., Randich, S., Testi, L., \& Biazzo, K.\ 2011, A\&A, 525, A47
\bibitem[Robberto et al.(2004)]{robberto2004} Robberto, M., Song, J., Mora Carrillo, G., et al.\ 2004, ApJ, 606, 952
\bibitem[Romaniello et al.(2004)]{romaniello2004} Romaniello, M., Robberto, M., \& Panagia, N.\ 2004, ApJ, 608, 220
\bibitem[Sergison et al.(2013)]{sergison2013} Sergison, D.~J., Mayne, N.~J., Naylor, T., Jeffries, R.~D., \& Bell, C.~P.~M.\ 2013, arXiv:1306.2282
\bibitem[Sicilia-Aguilar et al.(2010)]{sicilia-aguilar2010} Sicilia-Aguilar, A., Henning, T., \& Hartmann, L.~W.\ 2010, ApJ, 710, 597
\bibitem[Sicilia-Aguilar et al.(2006)]{sicilia-aguilar2006} Sicilia-Aguilar, A., Hartmann, L.~W., F{\"u}r{\'e}sz, G., et al.\ 2006, AJ, 132, 2135
\bibitem[Siess et al.(2000)]{siess2000} Siess, L., Dufour, E., \& Forestini, M.\ 2000, A\&A, 358, 593
\bibitem[Shu et al.(1994)]{shu1994} Shu, F., Najita, J., Ostriker, E., et al.\ 1994, ApJ, 429, 781
\bibitem[Spezzi et al.(2012)]{spezzi2012} Spezzi, L., de Marchi, G., Panagia, N., Sicilia-Aguilar, A., \& Ercolano, B.\ 2012, MNRAS, 421, 78
\bibitem[Tan et al.(2006)]{tan2006} Tan, J.~C., Krumholz, M.~R., \& McKee, C.~F.\ 2006, ApJL, 641, L121
\bibitem[Tannirkulam et al.(2008)]{tannirkulam2008} Tannirkulam, A., Monnier, J.~D., Millan-Gabet, R., et al.\ 2008, ApJL, 677, L51
\bibitem[Tout et al.(1999)]{tout1999} Tout, C.~A., Livio, M., \& Bonnell, I.~A.\ 1999, MNRAS, 310, 360
\bibitem[Tognelli et al.(2011)]{tognelli2011} Tognelli, E., Prada Moroni, P.~G., \& Degl'Innocenti, S.\ 2011, A\&A, 533, A109
\bibitem[Tognelli et al.(2013 \emph{in prep.})]{tognelli2013} Tognelli, E., Prada Moroni, P.~G., \& Da Rio, N.\ 2011, \emph{in prep.}
\bibitem[Valenti et al.(1993)]{valenti1993} Valenti, J.~A., Basri, G., \& Johns, C.~M.\ 1993, AJ, 106, 2024
\bibitem[Zhu et al.(2009)]{zhu2009} Zhu, Z., Hartmann, L.,  Gammie, C., \& McKinney, J.~C.\ 2009, ApJ, 701, 620
\end{thebibliography}
\end{document}